\documentclass[12pt]{article}

\usepackage[hypertexnames=false, naturalnames=false, dvips, colorlinks=true, linkcolor=black, citecolor=black, hidelinks]{hyperref}
\usepackage{cleveref}

\usepackage{url}

\usepackage{amsmath}
\usepackage{amssymb}

\usepackage{graphicx}

\usepackage{cite}

\usepackage{color} 

\usepackage[labelfont=bf,labelsep=period,justification=raggedright]{caption}

\makeatletter
\renewcommand{\@biblabel}[1]{\quad#1.}
\makeatother

\date{}

\newcommand{\noi}{{\noindent}}

\newlength\halfpage
\newlength\quartpage
\newlength\sixthpage
\newlength\mutation
\newlength\mutscheme
\usepackage{array}
\newcolumntype{x}[1]{>{\centering\arraybackslash\hspace{0pt}}p{#1}}

\usepackage{figuresupp}
\usepackage{tablesupp}

\usepackage{notes2bib}

\begin{document}

\begin{flushleft}
{\Large
\textbf{\emph{In silico} prediction of mutant HIV-1 proteases  cleaving a target sequence}
}
%
\\
\vspace{0.5cm}
Jan H. Jensen,$^{1}$
Martin Willemo{\"{e}}s,$^{2}$
Jakob R. Winther,$^{2}$
Luca De Vico$^{1,\ast}$
\\
\vspace{0.5cm}
\textbf{1} Department of Chemistry, University of Copenhagen,
Universitetsparken 5, DK-2100 Copenhagen, Denmark
\\
\textbf{2} Department of Biology, University of Copenhagen,
Ole Maal{\o}es Vej 5, DK-2200 Copenhagen, Denmark
\\
$\ast$ Corresponding Author, Email: luca@chem.ku.dk
\end{flushleft}

\section*{Abstract}

HIV-1 protease represents an appealing system for directed enzyme re-design, since it has various different endogenous targets, a relatively simple structure and it is well studied.
Recently Chaudhury and Gray (Structure (2009) 17: 1636 -- 1648) published a computational algorithm to discern the specificity determining residues of HIV-1 protease.
    In this paper we present two computational tools aimed at re-designing HIV-1 protease, derived from the algorithm of Chaudhuri and Gray. First, we present an energy-only based methodology to discriminate cleavable and non cleavable peptides for HIV-1 proteases, both wild type and mutant. 
Secondly, we show an algorithm we developed to predict mutant HIV-1 proteases capable of cleaving a new target substrate peptide, different from the natural targets of HIV-1 protease.
The obtained \emph{in silico} mutant enzymes were analyzed in terms of cleavability and specificity towards the target peptide using the energy-only methodology. We found two mutant proteases as best candidates for specificity and cleavability towards the target sequence.


\section*{List of Abbreviations}

\begin{description}
\item[PR] HIV-1 protease.
\item[WT-PR] Wild type HIV-1 protease.
\item[mutant PR] Mutant HIV-1 protease.
\item[Pr3 set] Set of mutant proteases derived from Pr3, a mutant protease developed by Alvizo \emph{et al}. These were heterodimer proteases.
\item[DR set] Set of mutant proteases derived as a subset of HIV-1 proteases that have been found to be drug resistant. These were homodimer proteases.
\end{description}

\section*{Introduction}

Proteases represent a class of enzymes ubiquitous in all living organisms, with multiple applications in industry and biotechnology research \cite{Rao1998, Gupta2002, Li2013}.
There is thus interest in designing new proteases capable of cleaving specific peptide sequences \cite{Beilen2002}. HIV-1 protease (PR) represents an attractive starting structure for directed enzyme re-design, since it is known to cleave a variety of sequences.
PR is the enzyme responsible for processing the gag -- pol fusion polyproteins of the HIV virus \cite{Wong2003}. PR is an aspartic protease \cite{Navia1989, Wlodawer1989, Davies1990} and is a homodimer where each chain is composed of 99 residues. Wild type PR (WT-PR) is very specific for the endogenous cleavage sequences of the polyprotein (endogenous substrate peptides, Table~\ref{tablepeptides1}),
even if the source of this specificity is still not completely clear. 
A series of other non-endogenous peptides have also been found to be cleaved by PR.
The latest hypothesis on the origin of this specificity, called dynamic substrate envelope \cite{Jeyabalan2002, Ozen2011}, states that peptides fitting into the protease cavity through a certain number of hydrogen bonds will be bound and possibly cleaved nearly regardless of their amino acid composition. In fact, there is no clear trend in amino acid sequence (e.g. a negatively charged amino acid in position P1 or a hydrophobic one in position P2'). This suggests that with few mutations PR could be made to cleave other target peptide sequences
in a specific manner.

Many computational studies on PR, both wild type (WT) and drug resistant  mutant enzymes, are aimed at elucidating the affinity of the enzymes towards endogenous substrates and inhibitors to be used as drug candidates \cite{Wang2001, Sherman2008, Perez2010, Lemmon2012}.
Recently Chaudhury and Gray \cite{Chaudhury2009} published a computational algorithm specifically tailored for PR and aimed at the identification of the specificity determining residues. The algorithm is based on PyRosetta \cite{Chaudhury2010}, a python script-based interface to Rosetta \cite{Kaufmann2010}. Thanks to the algorithm the authors were able to predict accurate protease -- substrate complex structures (within 1.1 {\AA} rms of the corresponding crystal structure) and introduced an energetic discrimination of cleavable peptides.
More recently Alvizo \emph{et al.} \cite{Alvizo2012} employed computational methods to re-engineer a mutant PR (Pr3) more specific for one of the endogenous peptide sequences over two others. 

The first aim of this study is to develop an energy-only based methodology to discern cleavable and non cleavable peptides for PRs, WT and mutant. This methodology is based on the qualitative evaluation of PR:peptide complexes binding energies and is derived from the algorithm developed by Chaudhury and Gray.
The second aim is to search and define an algorithm to predict mutant PRs capable of cleaving a specific target peptide sequence different from any endogenous substrate. We use our cleavability discerning methodology on the suggested mutant proteases, in order to define the best guess in terms of specificity towards the peptide sequence. In other words, the sought after mutant structure has to show better and worse binding towards the target and endogenous peptides, respectively, than WT-PR. To the best of our knowledge ours is the first study aimed at predicting a mutant PR capable of cleaving  specifically a non endogenous peptide sequence.


The paper is organized as follows: first, we present our computed binding energies for known cleavable and non-cleavable peptides bound to WT-PR, selected peptides bound to a set of single, double and triple mutants (Pr3 set) derived from Pr3 as developed by Alvizo \emph{et al.}, and a set of known mutant PRs and peptide derived from drug resistance (DR set) studies \cite{Kolli2009, Ozen2012, Prabu2004}.
Secondly, we present two different versions of our algorithm to determine mutant PRs that will cleave the sequence HFLSF*MAIP, where the * symbol indicates the desired cleaving site. A discussion about the best strategy to suggest mutant enzymes follows. The conclusions summarize the main findings of the paper, followed by a detailed description of the employed computational methods.

\section*{Results and Discussion}

\subsection*{Development of a Cleavability Test}

In general, the activity of an enzyme towards two similar substrates is regulated by (i) the strength of the enzyme-substrate binding and (ii) the efficiency of the enzymatic reaction. 
The two processes are regulated by two constants, usually indicated as $k_{m}$ and $k_{cat}$, respectively. The overall enzymatic efficiency is given by the ratio of these two constants.
The dynamic substrate envelope hypothesis \cite{Ozen2011} suggests that if a peptide is bound to PR it will be cleaved. Thus, we decided to evaluate the binding energy of different peptides to PR, which can be correlated to $k_{m}$. We then compared the computed binding energies to PR of known cleavable and non cleavable peptides, to be correlated to corresponding ranges of binding energies. By so doing we disregarded $k_{cat}$, that is we did not consider possible effects from the enzymatic reaction. 


The cleavability test was developed by considering binding energies of WT-PR with its endogenous and known cleavable substrates and known non-cleavable peptides. Afterwards we investigated the reliability of the test with mutant PRs (the Pr3 set) when binding PR endogenous substrates. Finally, we assessed the test on mutant PRs (the DR set) when binding mutant substrates.

The complete methodology for evaluating binding energies is described in the computational methods section. In brief, it is composed by a 
structure
optimization algorithm, 
followed by an energetic re-evaluation of the obtained structures.
%
%
In the following paragraphs we evaluate our methodology in terms of binding energies versus cleavability for: (1) WT-PR and its endogenous substrates and known cleavable and non-cleavable peptides, (2) the Pr3 set of mutant PRs and endogenous substrates and (3) the DR set with wild type and mutated endogenous peptides. Binding energies were computed also for WT-PR and all mutant PRs in complex with octa-alanine (poly-Ala) and octa-arginine (poly-Arg) peptides to test for aspecific binding.

%
(1)
Table~\ref{tablebinding1} reports the computed binding energies of the set of known cleavable endogenous peptides of WT-PR. The sequence of the tested endogenous cleavable peptides is reported in Table~\ref{tablepeptides1}.
%
%
Alongside the endogenous peptides, a set of 59 known cleavable peptides was also tested. The sequence of the 59 tested non-endogenous cleavable peptides was obtained as previously described \cite{Chaudhury2009,
Toszer1991, Riviere1991, Oswald1991, Tomaszek1992, Tomasselli1993}. Table~\ref{tablebindingextra} reports the computed binding energies to WT-PR and Table~\ref{tableextrapeptides} the sequences of these non-endogenous peptides.
%
%
Table~\ref{tablebinding2} reports the computed binding energies of a set of  peptides supposedly non-cleavable by WT-PR. The sequence of the 43 tested non-cleavable peptides was obtained as previously described \cite{Chaudhury2009, Tomasselli1993, Chou1996} and is reported in Table~\ref{tablepeptides2}. 

%
%
%
%
We performed a Mann-Whitney's U test \cite{Mann1947} to compare the computed binding energies, and found a significant difference between the cleavable and non-cleavable sets (p $\approx 10^{-7}$), as reported in Table~\ref{tableenergiescomparison}. Thus, we deemed the binding energy criterion sufficient to achieve discrimination.
{We further analyzed the computed binding energies through an ROC plot \cite{Metz1978} relative to different cutoff values, so as to differentiate between cleavable and non-clevable peptides. The plot is reported in Figure~\ref{ROCtotal}, and the relative data in Table~\ref{tableROCdata}. The computed area under ROC \cite{Hanley1982} is 0.79 and 0.80 for FMO and RosettaDock energies, respectively, being the values of 0.50 and 1.00 typical correspondingly of a useless and a perfect test. Through the ROC plot, we found the best cutoff values discerning cleavable and non-cleavable peptides as those closest to (0, 1), which represents the theoretical perfect test. We found that cutoff values of -25 kcal/mol and -3 kT are best at discerning FMO and RosettaDock computed binding energies, respectively.
Both FMO and RosettaDock perform well in computing binding energies capable of discerning cleavable and non-cleavable peptides. However, Figure~\ref{correlation} shows that there is no apparent correlation between FMO and RosettaDock computed binding energies. Thus, we repeated the Mann-Whitney's U test and ROC analysis excluding the set of non-endogenous known cleavable peptides binding energies. The rationale behind this analysis is that we expect WT-PR to bind the endogenous peptides with higher affinity, as opposed to the broader range of the complete cleavable set, characterized only by cleavability and not specificity. Consequently, we assume that the endogenous peptides set have better binding energies, than the complete set of cleavable peptides. The Mann-Whitney's U test (Table~\ref{tableenergiescomparison}) shows that the RosettaDock based binding energies are in this case two orders of magnitude worse than FMO at discerning cleavable and non-cleavable peptides. The relative ROC plot (Figure~\ref{ROCendogenous}) shows as well that the FMO data performs better than RosettaDock, in terms of more strict best cutoff value and larger area under the ROC. Thus, we concluded that FMO computed binding energies are better than RosettaDock ones since are capable of discerning expected effects, such as the usage of a better performing subset of peptides. In the rest of this paper we will discuss only binding energies computed through FMO energy re-evaluation.
%
%

From Table~\ref{tablebinding1} it is expected that WT-PR exhibits qualitatively different binding to the poly-protein substrates, given their computed binding energies ranging from -41 for the binding of p6pol-PR to -72 kcal/mol for p2-NC, with an average value of -60 kcal/mol. However, available experimental $K_m$ values\cite{Toszer1991} do not show any trend similar to the computed data.
Still, one has to remember that these computed binding energies should be considered only qualitatively and only compared to others obtained in the same manner. See the Computational Methods section for further details.
Furthermore, the span of both computed energies for which experimental data are available (20 kcal/mol) and the $K_m$ values (2 orders of magnitude) is too small to allow a clear trend. 
The computed binding energies for the set of cleavable non-endogenous peptides span a wide range of values, from -2 to -86 kcal/mol, with average -40 kcal/mol. These peptides not being the natural target of WT-PR may account for this large span. The average computed binding energy for all cleavable peptides is -43 kcal/mol.
The computed binding energies for the non-cleavable set of peptides (Table~\ref{tablebinding2}) span an even wider range of values than those of the cleavable ones. Some PR -- peptides complexes show positive energies. The majority (56\%) of the computed binding energies are in the range -35 -- 0 kcal/mol. However, a few peptides show a binding energy to WT-PR similar to those of the cleavable peptides.
%
%
%
%

(2)
Recently Alvizo \emph{et al.} \cite{Alvizo2012} suggested through computational means a triple mutant (Pr3) with increased binding capability towards the endogenous RTp51-RTp66 cleavage sequence peptide compared to that towards other two cleavage sequences CA-p2 and p2-NC. The efficiency of Pr3 in cleaving preferentially RTp51-RTp66 was later experimentally verified.
Pr3 was made by tethering a mutated chain of protease (A28S, D30F, G48R) to a wild type one. 
For comparison with our predicted mutant PRs, Table~\ref{tablebindingmutantspr3} reports our computed binding energies for the Pr3 three-fold mutant, as well for simpler one- and two-fold mutant PRs derived from Pr3 (Pr3 set), as compared to WT. Note, however, that experimental data are available only for the three-fold mutant PR. In our calculations, Pr3 set carried mutations only on chain A, while still being formed by two separate chains.
We expected to find that Pr3 computed binding would be stronger towards RTp51-RTp66, while weaker towards CA-p2 and p2-NC, compared to WT-PR.
The computed binding energies of the Pr3 set show that the mutant enzymes often have higher affinity for the desired RTp51-RTp66 peptide compared to CA-p2 and p2-NC. Most notably the double mutant A28S/G48R has a stronger computed binding energy towards the target peptide than WT-PR, while lower for the other two endogenous substrates.
%
The binding energy test indicates that A28S/G48R (for which there is no experimental data available) would have been a more successful mutation than Pr3. Nevertheless, the possibility of using the binding energy test with mutant PRs was found viable.

%
(3)
Finally we decided to apply the binding energy test to series of mutant PRs binding mutant endogenous substrates. Thus, we evaluated the binding energies of drug resistant HIV-1 proteases towards wild type and mutant substrate peptides. It has been found that mutations of the cleavage sites are correlated to mutations of the protease, often leading to drug resistance. We analyzed the K436R and A431V mutations of the NC-p1 Gag substrate peptide cleavage sequence in relation to a series of single mutations and one double mutation of HIV-1 protease (DR set). It has been reported \cite{Kolli2009} that a K436R mutation increases resistance to protease inhibitor drugs when combined with I50V, I84V and I84V/L90M PR mutations, while the A431V mutation results in a more efficient PR regardless of other mutations. We expected that the more efficient mutant PR -- mutant peptide combinations were also characterized by stronger binding energies.
Table~\ref{tablebindingresistance} reports the results of our binding energy test for the DR set.
%
Our methodology indicates cleavability for all combinations of mutant PRs and mutated NC-p1 substrate peptides. While there are some fluctuations in the binding energies, no clear pattern arises that can be related to the experimental findings. Possibly, the increased efficiency of drug resistant mutant proteases towards mutated peptides is 
related to $k_{cat}$. As previously stated, the effects of this constant are not considered by the present approach.
%
%
Nevertheless, the binding energy test was found suitable also for combinations of mutant PRs with any peptide.


%


\subsection*{Prediction and Analysis of Mutant PRs}

The second aim of this study was to develop a computational methodology for the design of a mutant PR. The sought after enzyme had to be capable of cleaving a new target substrate different from the endogenous ones. The obtained mutant PR should also be specific for the target peptide sequence compared to the endogenous peptides.
%
The chosen sequence for the target peptide was HLSF*MAIP, where the * symbol indicates the desired cleaving site. The sequence was extracted from that of $\kappa$-casein.
%
Once candidate mutant PRs were obtained, we employed the binding energy test to asses the enzymes cleaving capabilities. 
The possibility of an increase in cleaving capability towards the target substrate was asserted by differences in binding energy between WT-PR and  mutant PRs.
We evaluated the binding energies of mutant PRs in complex with the TF-PR peptide, used as a starting template (see the Computational Methods section), and the CA-p2 and p2-NC peptides (for selected mutant PRs) in order to test the specificity of our mutant PRs.

%
The mutant-generating algorithm
is described in details in the Computational Methods section. Two main strategies (Strategy1 and Strategy2) were employed for generating mutant PRs. In Strategy1, the side chains of only the 6 residues previously indicated as specificity determining\cite{Chaudhury2009} were allowed to change. The analysis of the binding energies of the mutant PRs generated by Strategy1 found the enzymes insufficient to perform the desired scope. This prompted us to further develop the algorithm.
In Strategy2, the side chains of 26 residues were allowed to change. See the Computational Methods section for further details on the residues choice. The analysis of the binding energies of these mutant PRs found some of the predicted enzymes to be adequate to cleave the desired target sequence. 







Tables~\ref{tablemutants-str1} and \ref{tablebindingmutants} in the Supporting material reports the Strategy1 mutant PRs ({\bf M1} -- {\bf M16}) and their computed binding energies towards the targetpeptide and TF-PR, CA-p2 and p2-NC endogenous peptides.
Among these mutant PRs, {\bf M5} shows the strongest binding energy towards the target peptide. However it has to be noted that the computed binding energy of {\bf M5} towards the TF-PR peptide (used as a starting template for all mutant enzymes) is also stronger with respect to WT. Possibly {\bf M5} is simply a better generic binder. 
To verify this hypothesis we tested {\bf M5} as a binder also for other two endogenous peptide sequences, CA-p2 and p2-NC. Compared to WT-PR, {\bf M5} has weaker binding energy for the former peptide, but equal for the latter. In conclusion, {\bf M5} is not predicted to be more specific for the target sequence than for the endogenous peptides.
Moreover, {\bf M5} was not directly predicted through Strategy1, but as a homodimeric derivative of {\bf M2}, which shows only a small improvement in binding of the target peptide.
All other mutant PRs suggested by Strategy1, {\bf M1} -- {\bf M4} and {\bf M6} -- {\bf M16}, were found having a weak binding energy towards the target peptide, with some of them showing prominently positive binding energies.
%
%
It can be concluded that Strategy1 is unsatisfactory at predicting a mutant PR with an increased and specific affinity towards the target peptide. This is possibly due to the fact that allowing only six residues to change is too strict a condition to achieve a suitable mutant PR.

Thus, we decided to further improve the mutation algorithm by including more residues among those that can be changed. 
%
The six generations of mutant PRs computed through our Strategy2 mutant algorithm are presented in Table~\ref{tablemutants-str2}. 
We refer to them as generations since at each macro step of the algorithm the lowest in energy (as computed with the standard RosettaDock energy function) structure was used as starting point for the next step.
The sixth generation ({\bf M23}) did not produce any new change with respect to the fifth ({\bf M22}), and the algorithm was consequently terminated. For each generation the structure with the lowest absolute energy was further optimized. After generation 1 two mutant structures were chosen ({\bf M17} and {\bf M18}) since they are very close in energy (as evaluated with the RosettaDock energy function, data not shown) but relatively different as mutation sites. In addition, an extra mutant PR ({\bf M24}) was generated as homodimer of {\bf M22}. 
The computed binding energies of the Strategy2 mutant PRs ({\bf M17} -- {\bf M24}) are shown in Table~\ref{tablebindingmutantstr3}. All Strategy2 mutant PRs show a binding energy towards the target sequence two to four fold stronger than WT-PR, with {\bf M17} displaying the strongest binding energy. However, as for {\bf M5}, binding energies towards the template peptide TF-PR as well as CA-p2 and p2-NC are also stronger than WT. Possibly {\bf M17} is also a good but generic binder.
Through the subsequent generations of mutant proteases, at last {\bf M22} shows a binding energy towards the target peptide more than three fold stronger than WT, while the computed binding energy towards the natural endogenous substrates is weaker than WT. Similar results were obtained for its homodimer {\bf M24}. {\bf M22} and {\bf M24} show binding energies below the cutoff value of -25 kcal/mol, and thus represent the best candidates to be further studied experimentally.


%
We compared the structures of WT-PR and {\bf M24} as optimized while binding the target peptide. Figure~\ref{backbones} reports the superimposed backbones of the two enzymes after structure alignment. The two computed structures are quite coincident. Hence, it is expected that {\bf M24} should retain the main structural features of the wild type enzyme.
We also tried to analyze the choice of changed residues. Figure~\ref{mutated-residues} shows that the residues that were changed from WT-PR to {\bf M24} are disposed all around the bound peptide. Figures~\ref{mutations-chain-a-28} -- \ref{mutations-chain-b-82} given as supporting material compare each residue that differs between WT-PR and {\bf M24}, while bound to the target peptide. Although it is evident that the A28S substitution on chain A introduces a hydrogen bond between the residue and the side chain of the serine in the peptide (Figure~\ref{mutations-chain-a-28}), the other substitutions are less easily rationalized. On going study aims at elucidating the role of the other residues substitutions.

It is interesting to note that Strategy2 mutated only 7 out of the 26 residues that were set as mutable in the method. It is also worth noting that of the 7 residues (A28, D30, K45, I50, P81, V82, I84) suggested by Strategy2 in the various mutant generations,  A28, K45, P81 are not included in the set of major mutations site of HIV-1 protease responsible for drug resistance \cite{Johnson2013}, that is: D30, V32, M46, I47, G48, I50, I54, Q58, T74, L76 V82, N83, I84, N88, L90.  A28, K45, and P81 together with I50 are also not included in the specificity determining residues set \cite{Chaudhury2009}. However, A28 was located by Alvizo \emph{et al}. for the Pr3 mutant \cite{Alvizo2012}.
We envision Strategy2 also as a tool to locate those residues most involved in binding a given substrate peptide.


From the analysis of the different PRs, mutant and wild type, and their binding energies, it is worth to note that WT-PR has a certain affinity with the octa-arginine peptide. Its computed binding energy is at the limit to consider the octa-arginine peptide as cleavable by WT-PR. Possibly this relatively strong binding is given by very few interactions. Accordingly, the single D30F change on chain A, that is changing one negatively charged residue into an aromatic hydrophobic one, is able to drop the computed binding energy to 0, as shown in Table~\ref{tablebindingresistance}. The currently going analysis of the residue by residue interactions for the modified side chains will give further information also on this aspect of the binding of PR.

Finally, it is interesting to note that the algorithm is not always preserving amino acid side chain changes through the generations. For example, I84V on chain A is introduced in {\bf M18} and kept in {\bf M19}, {\bf M20} and {\bf M21}, but later reverted. Possibly, an isoleucin in position 84 is energetically more favorable, given the other side chain changes.

\section*{Conclusions}


In the first part of this study we developed a methodology to test the cleavability of a peptide by HIV-1 protease (Tables~\ref{tablebinding1} and \ref{tablebinding2}), solely based on the binding energy between the enzyme and the substrate. The methodology can also be applied to mutant PRs, Table~\ref{tablebindingmutantspr3}. The technique is based on a PyRosetta algorithm generating, iteratively, optimized structures, coupled with an energy re-evaluation at a higher level of theory (FMO/PCM MP2/6-31G(d)).



In the second part of this study, the
optimization algorithm was extended to permit the stochastic change of the side chain of selected residues, in order to better bind a given target peptide sequence. The selected target peptide was required to be different from the endogenous peptides. The desired outcome was a mutant PR with stronger and weaker predicted binding energy for the target and endogenous peptides, respectively, compared to WT-PR. The mutant PRs {\bf M22} and {\bf M24} generated through Strategy2 exhibit such desired characteristics (Table~\ref{tablebindingmutantstr3}). 
%
%
We analyzed the backbone structure of WT-PR and {\bf M24} and found no major differences, thus indicating that {\bf M24} should retain the general structure features of wild type HIV-1 protease. 
Strategy2 algorithm is able to predict mutations outside the usual set of residues involved in drug resistance, possibly giving an ulterior insight into the binding process of HIV-1 protease.


Ongoing experimental studies will show if and how well {\bf M22} and/or {\bf M24} bind and cleave the target sequence. Our current experimental and computational studies are also aimed at analyzing {\bf M24} mutations, residue by residue and in combination, and their possible role in binding the target sequence.
It is our hope that the experimental tests will provide enough information to be used  to further improve the mutant generating algorithm.
If the combination of computational algorithm and experimental verification is successful it will maybe permit the design of mutant PRs specific for any given substrate peptide.

\section*{Computational Methods}

In general, the activity of an enzyme towards two similar substrates is regulated by (i) how good the enzyme-substrate binding is and (ii) how efficient the enzymatic reaction is. Following the dynamic substrate envelope hypothesis \cite{Jeyabalan2002, Ozen2011}, we assume a correlation between the binding of different peptides to PR and cleavability of the former. Thus we compute qualitative binding energies, on the premise that lower binding energy equals better cleavability.

\subsection*{Binding Energies}

\subsubsection*{PyRosetta Algorithm}

The structure of wild type (WT) HIV-1 protease in complex with different octa-peptides was optimized using PyRosetta 1.1,\cite{Chaudhury2010} a python script-based interface to Rosetta,\cite{Kaufmann2010} and the algorithm depicted in Figure~\ref{flowchart1}. The algorithm is based on the flexible peptide-docking algorithm used by Chaudhury and Gray\cite{Chaudhury2009} to identify in WT HIV-1 protease the active-site residues mostly involved in the discrimination of cleavable and non-cleavable peptides. Following their algorithm, the HIV-1 protease -- peptide complexes are represented in atomic resolution, as opposed to a coarse-grain representation. With respect to the algorithm described in \cite{Chaudhury2009}, our algorithm (Figure~\ref{flowchart1}) has a larger number of cycles (8x4x6=192 compared to 8x12=96), and more 'small' and 'shear' moves for the perturbation of both the side chain and the backbone atoms. The side chain conformations are further optimized through a repacking algorithm\cite{Kuhlman2000} and using the extended Dunbrack library\cite{Dunbrack1997, Wang2005}. The moves are applied to all residues of the substrate peptide plus a selected number of residues of the protease, with the following criterion: all residues inside a 5 {\AA} distance from any atom of the substrate peptide, plus all the residues reported as active by Chaudhury and Gray\cite{Chaudhury2009}, plus their $\pm$1 neighbours, plus if one residue is included on only one chain it is made to be included in both. After the moves, an energy minimization step is performed, based on the Davidon-Fletcher-Powell method \cite{davidon1991, Fletcher1963}. Each structure is then accepted or rejected based on a Monte Carlo (MC) criterion depending on the standard RosettaDock energy function \cite{Gray2003, Dunbrack1997, Kuhlman2000, Lazaridis1990, Kortemme2002}. Along the optimization a temperature gradient was applied, from an initial value of kT = 3.0 to 1.0, unless differently stated. 500 decoy structures were generated using 5 parallel algorithm runs, each producing 100 structures.

The main difference with the algorithm of \cite{Chaudhury2009} is that after the algorithm produced 500 decoy structures, the lowest in energy is chosen and used as a starting structure for another cycle of optimization. This process is repeated $K$ times, until convergence. It was found that, after at least 5 cycles, the computed RosettaDock energy did not change between subsequent cycles as soon as all 5 parallel runs of a single cycle produced structures with the same energy. Consequently, in order to render as automatic as possible the algorithm, the fact that $K > 5$ and that each parallel run produced, as best structure, a decoy with the same energy was taken as a mark for convergence. It was found that, on average, a value of $K = 20$ was sufficient. As an example, Figure~\ref{convergence} reports the energy of WT-PR bound to TF-PR along the optimization. The points at each step corresponds to the RosettaDock energy of the lowest in energy decoy out of the 500 computed at that particular step. Such structure would then be used as starting point for the next cycle. At the end of the $K$ cycles the lowest in energy decoy is chosen as the PyRosetta optimized structure.


The same algorithm was also used for the optimization of mutant HIV-1 proteases (\emph{vide infra}), the octa-peptides alone, and the protease alone as apo-protein.

The starting structures were prepared from that of HIV-1 protease in complex with an inhibitor (PDB accession code 1HXB \cite{Jaskolski1991}), considered as apo-protein. In order to place the substrate peptide, the structure of a D25N deactivated protease in complex with the natural substrate peptide p2-NC (PDB accession code 1KJ7 \cite{Jeyabalan2002}) was aligned with respect to the backbone atoms of the protease (RMS = 0.436 \AA). The starting structure was then composed using the apo-protein from 1HXB and the substrate peptide from 1KJ7. All subsequent protease-peptide complexes were created starting from this structure and mutating the peptide accordingly. See Table~\ref{tablepeptides1}, Table~\ref{tablepeptides2} and Table~\ref{tablepeptides3} for a complete list of the considered substrate peptides. Hydrogen atoms were added through the program Pymol \cite{pymol}.

\subsubsection*{Further Structures Optimization and Energetic Re-evaluation}


The position of the hydrogen atoms of each PyRosetta generated structure was optimized using Open Babel\cite{Boyle2011} with the MMFF94\cite{MMFF94-1,  MMFF94-2, MMFF94-3,  MMFF94-4, MMFF94-5} force field.
%
%
%
%
The energy of each structure was finally re-evaluated at the higher level of theory `FMO2-MP2/6-31G(d)/PCM[1]'. Single point energy evaluations were carried out using the fragment molecular orbital (FMO) approximation \cite{fmo2002, fmo2007}, as implemented in GAMESS \cite{Schmidt1993}.
Each FMO calculation was carried out at the MP2 level of theory \cite{Fedorov2004} with the 6-31G(d) basis set \cite{Hariharan1973, Francl1982} and the Polarazible Continuum Model (PCM) approximation \cite{Tomasi2005, Fedorov2006}.
Pairs of fragments separated by more than two van der Waals radii were calculated using a Coulomb expression for the interaction energy and ignoring correlation effects (RESDIM=2.0 RCORSD=2.0 in \$FMO).
The input files for the FMO calculations were prepared using the program FRAGIT \cite{Steinmann2012}.

\subsubsection*{Binding Energies Evaluation}

The re-evaluated energy of every optimized structure was used to compute the binding energy of PR with different substrate peptides. The binding energy ($E_{Bind}$) of HIV-1 protease (wild type or mutated) and a peptide was evaluated with equation~\eqref{binden}, where $E_{Complex}$ is the energy of the complex, $E_{APO}$ the energy of the protease optimized as apo-protein, $E_{Pep}$ the energy of the optimized peptide.

\begin{equation}
E_{Bind} = E_{Complex} - \left( E_{APO} + E_{Pep} \right)
\label{binden}
\end{equation}

These binding energies can not be directly compared to experimental values, for which a much more complex and accurate methodology is required \cite{Genheden2010}. These energies were used only to qualitatively compare different PR -- peptide combinations.

%
%

\subsection*{Mutation Algorithm}

A similar procedure as that described in Figure~\ref{flowchart1} was used to produce mutant HIV-1 proteases, possibly capable of cleaving a given peptide different from the endogenous substrate peptides. The general idea was to 'expose' the protease to a different peptide and allow some residues to change in order to accommodate it better. A target octa-peptide was chosen: HLSF*MAIP, where the * symbol indicates the desired cleaving site. The peptide sequence was extracted from that of $\kappa$-casein.

The assumption behind the algorithm is that lowering the energy of the PR -- peptide complex by changing the side chains of selected residues would decrease also the binding energy, thus increasing the cleavability.

Two different methodologies were designed to predict mutant PRs, Strategy1 and Strategy2. The Strategy1 mutation algorithm is depicted in Figure~\ref{flowchart2}. Each optimization step corresponds to the algorithm of Figure~\ref{flowchart1}. In the mutation steps (also based on the previous algorithm), the Dunbrack library of rotamers includes all rotamers of all amino acids, but only for a selected number of residues. The six specificity determining residues, as found by \cite{Chaudhury2009}, are chosen to be altered.
In other words, during the mutation step, whenever one of the alterable residues is being optimized, the random choice of a test rotamer is among all possible amino acids.
In Scheme A alterations are allowed on all 6 residues on both chains, for a total of 12 alterable residues. Thus, side-chain perturbation and repacking rotamer choice is performed randomly selecting among 12 x 20 = 240 possible amino acids.
In Scheme B only alterations on L76 and V82 of Chain A and D30, I47, G48, and I84 of Chain B are allowed, for a total of 6 alterable residues. In this case, side-chain perturbation and repacking rotamer choice is performed with a random selection among 6 x 20 = 120 possible amino acids.
%
Each mutation step took ca. 40 hours on 5 cpus to produce 500 decoys. The lowest energy decoy is then chosen as starting structure for the next step. The energy of the structure is evaluated with the standard RosettaDock energy function. The residue reference energy part of the energy function \cite{Kuhlman2000} takes into account also the differences between different amino acids. In other words, energy differences between two mutant structures originates solely from different side chain interactions rather than from a different number of atoms.

Both the mutation and the optimization steps were repeated $K^{\prime}$ and $K$ times, respectively. The mutation cycles are considered converged once two following cycles do not introduce new mutations. Different values of $K$ and $K^{\prime}$ were found necessary to reach convergence. After a  series of mutation cycles ($K^{\prime}\ge8$), a series of optimization cycles was performed  ($K\ge8$), followed by another usually shorter mutation cycle ($K^{\prime}\le3$) and finally a short optimization cycle ($K\le3$).

Among the naturally cleaved peptides, TF-PR (sequence SFN\underline{F}*PQ\underline{I}T) was chosen as a starting substrate peptide, since it is the most similar, in terms of conserved residues, to the target peptide (sequence HLS\underline{F}*MA\underline{I}P). Consequently, the optimized structure of WT protease in complex with the TF-PR peptide was chosen as starting template. The substrate peptide sequence was altered one amino acid at the time, as reported in Table~\ref{tablemutation}. After each peptide alteration, a series of protease mutation and optimization cycles were performed. Once convergence was reached, a new peptide single amino acid change was introduced and the procedure repeated. Different mutant PRs were obtained from different runs by changing a few parameters, e.g. the initial temperature of the simulation. These parameters are specified in Table~\ref{tablemutants-str1}. Some mutant PRs were also produced by 'exposing' the protease directly to the target peptide without prior intermediates (mutation Scheme F). This last process required a higher number of $K^{\prime}$ cycles ($K^{\prime}\ge15$), but without having to cycle through one substrate peptide residue at the time.

All mutant PRs obtained through Strategy1 were heterodimers. By simply equalizing alterations on both chains a number of extra homodimer mutant PRs were also obtained. These structures were subsequently optimized as previously described.


In Strategy2 the number of residues allowed to change was increased in order to include all amino acids residing inside a 3 {\AA} radius from the TF-PR peptide. 
In other words, we chose those residues with at least one atom that is distant at most 3 {\AA} from any atom of the substrate peptide.
The specificity determining residues were also included in the set of alterable amino acids, if not already present. The residues Asp25, Thr26 and Gly27 of both chains were excluded from the set, since they represent the catalytic triad \cite{Wong2003}. The full set of 26 residues is reported in Table~\ref{tablemutstr2}.
Thus, side-chain perturbation and repacking rotamer choice is performed randomly selecting among 26 x 20 = 520 possible amino acids.
The mutant PRs were generated using the target peptide directly (Scheme F). Each mutation step took a bit more than 3 days on 5 cpus to produce 500 decoy structures. An initial temperature of 9 kT was usued. $K^{\prime}$ = 6 mutation cycles were performed. The lowest in energy decoy after each mutation step was subsequently optimized (two after the first step). The sixth mutation step did not introduce any new mutation in PR and the mutation cycle was stopped.

Also the mutant PRs obtained through Strategy2 were heterodimers. Only the homodimer of the final mutant PR was considered, see Table~\ref{tablemutants-str2}.


\section*{Acknowledgments}

Computational resources were provided by the Danish Center for Scientific Computing (DCSC).
LDV acknowledges S. Chaudhury and J. J. Gray for fruitful discussions about PyRosetta, for providing the set of non cleavable peptides and a copy of their algorithm script. C. Steinmann is acknowledged for help with the program FRAGIT and the FMO based calculations.

\section*{Supporting material}

Supporting material available:  Tables~\ref{tablepeptides1} -- \ref{tableROCdata}, Figures~\ref{correlation} -- \ref{mutations-chain-b-82}, a movie showing the three dimensional structure of WT-PR bound to the target peptide, with highlighted the residues that are changed in {\bf M24}.

\bibliography{biblio}

\begin{thebibliography}{10}
\providecommand{\url}[1]{\texttt{#1}}
\providecommand{\urlprefix}{URL }
\expandafter\ifx\csname urlstyle\endcsname\relax
  \providecommand{\doi}[1]{doi:\discretionary{}{}{}#1}\else
  \providecommand{\doi}{doi:\discretionary{}{}{}\begingroup
  \urlstyle{rm}\Url}\fi
\providecommand{\bibAnnoteFile}[1]{%
  \IfFileExists{#1}{\begin{quotation}\noindent\textsc{Key:} #1\\
  \textsc{Annotation:}\ \input{#1}\end{quotation}}{}}
\providecommand{\bibAnnote}[2]{%
  \begin{quotation}\noindent\textsc{Key:} #1\\
  \textsc{Annotation:}\ #2\end{quotation}}
\providecommand{\eprint}[2][]{\url{#2}}

\bibitem{Rao1998}
Rao MB, Tanksale AM, Ghatge MS, Deshpande VV (1998) {Molecular and
  Biotechnological Aspects of Microbial Proteases}.
\newblock Microbiology and Molecular Biology Reviews 62: 597-635.
\bibAnnoteFile{Rao1998}

\bibitem{Gupta2002}
Gupta R, Beg Q, Lorenz P (2002) {Bacterial alkaline proteases: molecular
  approaches and industrial applications}.
\newblock Applied Microbiology and Biotechnology 59: 15-32.
\bibAnnoteFile{Gupta2002}

\bibitem{Li2013}
Li Q, Yi L, Marek P, Iverson BL (2013) {Commercial proteases: Present and
  future}.
\newblock FEBS Letters 587: 1155 - 1163.
\bibAnnoteFile{Li2013}

\bibitem{Beilen2002}
van Beilen JB, Li Z (2002) {Enzyme technology: an overview}.
\newblock {Current Opinion in Biotechnology} 13: 338 - 344.
\bibAnnoteFile{Beilen2002}

\bibitem{Wong2003}
Brik A, Wong CH (2003) {HIV-1 protease: mechanism and drug discovery}.
\newblock Org Biomol Chem 1: 5-14.
\bibAnnoteFile{Wong2003}

\bibitem{Navia1989}
Navia MA, Fitzgerald PMD, {McKeever} BM, Leu CT, Heimbach JC, et~al. (1989)
  {Three-dimensional structure of aspartyl protease from human immunodeficiency
  virus HIV-1}.
\newblock Nature 337: 615-620.
\bibAnnoteFile{Navia1989}

\bibitem{Wlodawer1989}
Wlodawer A, Miller M, Jaskolski M, Sathyanarayana B, Baldwin E, et~al. (1989)
  {Conserved folding in retroviral proteases: crystal structure of a synthetic
  HIV-1 protease}.
\newblock Science 245: 616-621.
\bibAnnoteFile{Wlodawer1989}

\bibitem{Davies1990}
Davies DR (1990) {The Structure and Function of the Aspartic Proteinases}.
\newblock {Annual Review of Biophysics and Biophysical Chemistry} 19: 189-215.
\bibAnnoteFile{Davies1990}

\bibitem{Jeyabalan2002}
Prabu-Jeyabalan M, Nalivaika E, Schiffer CA (2002) {Substrate Shape Determines
  Specificity of Recognition for HIV-1 Protease: Analysis of Crystal Structures
  of Six Substrate Complexes}.
\newblock Structure 10: 369 - 381.
\bibAnnoteFile{Jeyabalan2002}

\bibitem{Ozen2011}
{\"{O}}zen A, Halilo{\u{g}}lu T, Schiffer CA (2011) {Dynamics of Preferential
  Substrate Recognition in HIV-1 Protease: Redefining the Substrate Envelope}.
\newblock {Journal of Molecular Biology} 410: 726 - 744.
\bibAnnoteFile{Ozen2011}

\bibitem{Wang2001}
Wang W, Kollman PA (2001) {Computational study of protein specificity: The
  molecular basis of HIV-1 protease drug resistance}.
\newblock Proceedings of the National Academy of Sciences 98: 14937-14942.
\bibAnnoteFile{Wang2001}

\bibitem{Sherman2008}
Sherman W, Tidor B (2008) {Novel Method for Probing the Specificity Binding
  Profile of Ligands: Applications to HIV Protease}.
\newblock Chemical Biology {\&} Drug Design 71: 387--407.
\bibAnnoteFile{Sherman2008}

\bibitem{Perez2010}
Perez MAS, Fernandes PA, Ramos MJ (2010) {Substrate Recognition in HIV-1
  Protease: A Computational Study}.
\newblock The Journal of Physical Chemistry B 114: 2525-2532.
\bibAnnoteFile{Perez2010}

\bibitem{Lemmon2012}
Lemmon G, Kaufmann K, Meiler J (2012) {Prediction of HIV-1 Protease/Inhibitor
  Affinity using RosettaLigand}.
\newblock Chemical Biology {\&} Drug Design 79: 888--896.
\bibAnnoteFile{Lemmon2012}

\bibitem{Chaudhury2009}
Chaudhury S, Gray JJ (2009) {Identification of Structural Mechanisms of HIV-1
  Protease Specificity Using Computational Peptide Docking: Implications for
  Drug Resistance}.
\newblock Structure 17: 1636 - 1648.
\bibAnnoteFile{Chaudhury2009}

\bibitem{Chaudhury2010}
Chaudhury S, Lyskov S, Gray JJ (2010) {PyRosetta}: a script-based interface for
  implementing molecular modeling algorithms using {Rosetta}.
\newblock Bioinformatics 26: 689-691.
\bibAnnoteFile{Chaudhury2010}

\bibitem{Kaufmann2010}
Kaufmann KW, Lemmon GH, DeLuca SL, Sheehan JH, Meiler J (2010) {Practically
  Useful: What the Rosetta Protein Modeling Suite Can Do for You}.
\newblock Biochemistry 49: 2987-2998.
\bibAnnoteFile{Kaufmann2010}

\bibitem{Alvizo2012}
Alvizo O, Mittal S, Mayo SL, Schiffer CA (2012) {Structural, kinetic, and
  thermodynamic studies of specificity designed HIV-1 protease}.
\newblock Protein Science 21: 1029--1041.
\bibAnnoteFile{Alvizo2012}

\bibitem{Kolli2009}
Kolli M, Stawiski E, Chappey C, Schiffer CA (2009) {Human Immunodeficiency
  Virus Type 1 Protease-Correlated Cleavage Site Mutations Enhance Inhibitor
  Resistance}.
\newblock {Journal of Virology} 83: 11027-11042.
\bibAnnoteFile{Kolli2009}

\bibitem{Ozen2012}
{\"{O}}zen A, Halilo{\u{g}}lu T, Schiffer CA (2012) {HIV-1 Protease and
  Substrate Coevolution Validates the Substrate Envelope As the Substrate
  Recognition Pattern}.
\newblock {Journal of Chemical Theory and Computation} 8: 703-714.
\bibAnnoteFile{Ozen2012}

\bibitem{Prabu2004}
Prabu-Jeyabalan M, Nalivaika EA, King NM, Schiffer CA (2004) {Structural Basis
  for Coevolution of a Human Immunodeficiency Virus Type 1 Nucleocapsid-p1
  Cleavage Site with a V82A Drug-Resistant Mutation in Viral Protease}.
\newblock Journal of Virology 78: 12446-12454.
\bibAnnoteFile{Prabu2004}

\bibitem{Toszer1991}
T\"{o}zs\'{e}r J, Bl\'{a}ha I, Copeland TD, Wondrak EM, Oroszlan S (1991)
  {Comparison of the HIV-1 and HIV-2 proteinases using oligopeptide substrates
  representing cleavage sites in Gag and Gag-Pol polyproteins}.
\newblock {FEBS Letters} 281: 77 - 80.
\bibAnnoteFile{Toszer1991}

\bibitem{Riviere1991}
Rivi\`{e}re Y, Blank V, Kourilsky P, Isra\"{e}l A (1991) {Processing of the
  precursor of NF-$\kappa$B by the HIV-1 protease during acute infection}.
\newblock {Nature} 350: 625 - 626.
\bibAnnoteFile{Riviere1991}

\bibitem{Oswald1991}
Oswald M, von~der Helm K (1991) {Fibronectin is a non-viral substrate for the
  HIV proteinase}.
\newblock {FEBS Letters} 292: 298 - 300.
\bibAnnoteFile{Oswald1991}

\bibitem{Tomaszek1992}
Tomaszek TA, Moore ML, Strickler JE, Sanchez RL, Dixon JS, et~al. (1992)
  {Proteolysis of an active site peptide of lactate dehydrogenase by human
  immunodeficiency virus type 1 protease}.
\newblock Biochemistry 31: 10153-10168.
\bibAnnoteFile{Tomaszek1992}

\bibitem{Tomasselli1993}
Tomasselli AG, Sarcich JL, Barrett LJ, Reardon IM, Howe WJ, et~al. (1993)
  {Human immunodeficiency virus type-1 reverse transcriptase and ribonuclease h
  as substrates of the viral protease}.
\newblock Protein Science 2: 2167--2176.
\bibAnnoteFile{Tomasselli1993}

\bibitem{Chou1996}
Chou KC (1996) {Prediction of Human Immunodeficiency Virus Protease Cleavage
  Sites in Proteins}.
\newblock Analytical Biochemistry 233: 1 - 14.
\bibAnnoteFile{Chou1996}

\bibitem{Mann1947}
Mann HB, Whitney DR (1947) {On a Test of Whether One of Two Random Variables is
  Stochastically Larger than the Other}.
\newblock The Annals of Mathematical Statistics 18: 50--60.
\bibAnnoteFile{Mann1947}

\bibitem{Metz1978}
Metz CE (1978) {Basic principles of \{ROC\} analysis}.
\newblock {Seminars in Nuclear Medicine} 8: 283 - 298.
\bibAnnoteFile{Metz1978}

\bibitem{Hanley1982}
Hanley JA, McNeil BJ (1982) {The meaning and use of the area under a receiver
  operating characteristic (ROC) curve.}
\newblock Radiology 143: 29-36.
\bibAnnoteFile{Hanley1982}

\bibitem{Johnson2013}
Johnson VA, Calvez V, G{\"u}nthard HF, Paredes R, Pillay D, et~al. (2013)
  {Update of the Drug Resistance Mutations in HIV-1: March 2013}.
\newblock {Top Antivir Med} 21: 6--14.
\bibAnnoteFile{Johnson2013}

\bibitem{Kuhlman2000}
Kuhlman B, Baker D (2000) {Native protein sequences are close to optimal for
  their structures}.
\newblock Proceedings of the National Academy of Sciences of the United States
  of America 97: 10383-10388.
\bibAnnoteFile{Kuhlman2000}

\bibitem{Dunbrack1997}
Dunbrack RL, Cohen FE (1997) Bayesian statistical analysis of protein
  side-chain rotamer preferences.
\newblock Protein Science 6: 1661--1681.
\bibAnnoteFile{Dunbrack1997}

\bibitem{Wang2005}
Wang C, Schueler-Furman O, Baker D (2005) Improved side-chain modeling for
  protein-protein docking.
\newblock Protein Science 14: 1328--1339.
\bibAnnoteFile{Wang2005}

\bibitem{davidon1991}
Davidon WC (1991) {Variable Metric Method for Minimization}.
\newblock SIAM Journal on Optimization 1: 1-17.
\bibAnnoteFile{davidon1991}

\bibitem{Fletcher1963}
Fletcher R, Powell MJD (1963) {A Rapidly Convergent Descent Method for
  Minimization}.
\newblock The Computer Journal 6: 163-168.
\bibAnnoteFile{Fletcher1963}

\bibitem{Gray2003}
Gray JJ, Moughon S, Wang C, Schueler-Furman O, Kuhlman B, et~al. (2003)
  {Protein-Protein Docking with Simultaneous Optimization of Rigid-body
  Displacement and Side-chain Conformations}.
\newblock Journal of Molecular Biology 331: 281 - 299.
\bibAnnoteFile{Gray2003}

\bibitem{Lazaridis1990}
Lazaridis T, Karplus M (1999) {Effective energy function for proteins in
  solution}.
\newblock Proteins: Structure, Function, and Bioinformatics 35: 133--152.
\bibAnnoteFile{Lazaridis1990}

\bibitem{Kortemme2002}
Kortemme T, Baker D (2002) {A simple physical model for binding energy hot
  spots in protein�--protein complexes}.
\newblock Proceedings of the National Academy of Sciences of the United States
  of America 99: 14116-14121.
\bibAnnoteFile{Kortemme2002}

\bibitem{Jaskolski1991}
Jaskolski M, Tomasselli AG, Sawyer TK, Staples DG, Heinrikson RL, et~al. (1991)
  {Structure at 2.5-.ANG. resolution of chemically synthesized Human
  Immunodeficiency Virus Type 1 protease complexed with a hydroxyethylene-based
  inhibitor}.
\newblock Biochemistry 30: 1600-1609.
\bibAnnoteFile{Jaskolski1991}

\bibitem{pymol}
{The PyMol Molecular Graphics System, Version 1.2r1 Schr{\"o}dinger, LLC}.
\bibAnnoteFile{pymol}

\bibitem{Boyle2011}
O'Boyle N, Banck M, James C, Morley C, Vandermeersch T, et~al. (2011) {Open
  {Babel}: {An} open chemical toolbox}.
\newblock Journal of Cheminformatics 3: 33.
\bibAnnoteFile{Boyle2011}

\bibitem{MMFF94-1}
Halgren TA (1996) {Merck molecular force field. {I}. {Basis}, form, scope,
  parameterization, and performance of {MMFF94}}.
\newblock Journal of Computational Chemistry 17: 490--519.
\bibAnnoteFile{MMFF94-1}

\bibitem{MMFF94-2}
Halgren TA (1996) {Merck molecular force field. {II}. {MMFF94} van der {Waals}
  and electrostatic parameters for intermolecular interactions}.
\newblock Journal of Computational Chemistry 17: 520--552.
\bibAnnoteFile{MMFF94-2}

\bibitem{MMFF94-3}
Halgren TA (1996) {Merck molecular force field. {III}. {Molecular} geometries
  and vibrational frequencies for {MMFF94}}.
\newblock Journal of Computational Chemistry 17: 553--586.
\bibAnnoteFile{MMFF94-3}

\bibitem{MMFF94-4}
Halgren TA, Nachbar RB (1996) {Merck molecular force field. {IV}.
  {Conformational} energies and geometries for {MMFF94}}.
\newblock Journal of Computational Chemistry 17: 587--615.
\bibAnnoteFile{MMFF94-4}

\bibitem{MMFF94-5}
Halgren TA (1996) {Merck molecular force field. {V}. {Extension} of {MMFF94}
  using experimental data, additional computational data, and empirical rules}.
\newblock Journal of Computational Chemistry 17: 616--641.
\bibAnnoteFile{MMFF94-5}

\bibitem{fmo2002}
Nakano T, Kaminuma T, Sato T, Fukuzawa K, Akiyama Y, et~al. (2002) {Fragment
  molecular orbital method: use of approximate electrostatic potential}.
\newblock Chemical Physics Letters 351: 475 - 480.
\bibAnnoteFile{fmo2002}

\bibitem{fmo2007}
Fedorov DG, Kitaura K (2007) {Extending the Power of Quantum Chemistry to Large
  Systems with the Fragment Molecular Orbital Method}.
\newblock The Journal of Physical Chemistry A 111: 6904-6914.
\bibAnnoteFile{fmo2007}

\bibitem{Schmidt1993}
Schmidt MW, Baldridge KK, Boatz JA, Elbert ST, Gordon MS, et~al. (1993)
  {General atomic and molecular electronic structure system}.
\newblock Journal of Computational Chemistry 14: 1347--1363.
\bibAnnoteFile{Schmidt1993}

\bibitem{Fedorov2004}
Fedorov DG, Kitaura K (2004) {Second order M{\o}ller-Plesset perturbation
  theory based upon the fragment molecular orbital method}.
\newblock The Journal of Chemical Physics 121: 2483-2490.
\bibAnnoteFile{Fedorov2004}

\bibitem{Hariharan1973}
Hariharan PC, Pople JA (1973) {The influence of polarization functions on
  molecular orbital hydrogenation energies}.
\newblock Theoretical Chemistry Accounts: Theory, Computation, and Modeling
  (Theoretica Chimica Acta) 28: 213-222.
\bibAnnoteFile{Hariharan1973}

\bibitem{Francl1982}
Francl MM, Pietro WJ, Hehre WJ, Binkley JS, Gordon MS, et~al. (1982)
  {Self-consistent molecular orbital methods. {XXIII}. {A} polarization-type
  basis set for second-row elements}.
\newblock The Journal of Chemical Physics 77: 3654-3665.
\bibAnnoteFile{Francl1982}

\bibitem{Tomasi2005}
Tomasi J, Mennucci B, Cammi R (2005) {Quantum Mechanical Continuum Solvation
  Models}.
\newblock Chemical Reviews 105: 2999-3094.
\bibAnnoteFile{Tomasi2005}

\bibitem{Fedorov2006}
Fedorov DG, Kitaura K, Li H, Jensen JH, Gordon MS (2006) {The polarizable
  continuum model (PCM) interfaced with the fragment molecular orbital method
  (FMO)}.
\newblock Journal of Computational Chemistry 27: 976--985.
\bibAnnoteFile{Fedorov2006}

\bibitem{Steinmann2012}
Steinmann C, Ibsen MW, Hansen AS, Jensen JH (2012) {FragIt: A Tool to Prepare
  Input Files for Fragment Based Quantum Chemical Calculations}.
\newblock {PLoS ONE} 7: e44480.
\bibAnnoteFile{Steinmann2012}

\bibitem{Genheden2010}
Genheden S, Kongsted J, S{\"{o}}derhjelm P, Ryde U (2010) {Nonpolar Solvation
  Free Energies of Protein--Ligand Complexes}.
\newblock Journal of Chemical Theory and Computation 6: 3558-3568.
\bibAnnoteFile{Genheden2010}

\end{thebibliography}

\newpage

\section*{Figure Legends}

\begin{figure}[!ht]
\begin{center}
\includegraphics{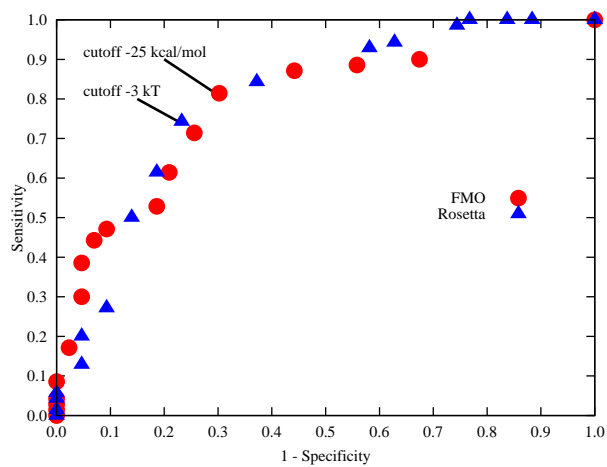}
\end{center}
\caption{
{\bf ROC plot comparing different cutoff values for binding energies computed through FMO energy re-evaluation or RosettaDock energy function.} The values for each method closest to the theoretical optimum (0,1) are highlighted. The computed area under the ROC curve is 0.79 and 0.80 for FMO and Rosetta, respectively. The raw data is reported in Table~\ref{tableROCdata}.
}
\label{ROCtotal}
\end{figure}

\begin{figure}[!ht]
\begin{center}
\includegraphics{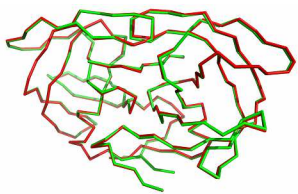}
\end{center}
\caption{
{\bf Backbone difference between PyRosetta computed structures of WT-PR and {\bf M24}.} The optimized structures of WT-PR and {\bf M24} binding the target peptide were aligned with respect to their $\alpha$-carbon atoms using PyMol. The backbone of {\bf M24} (red) is almost coincident with that of WT-PR (green) with a RMS of 0.227 \AA.
}
\label{backbones}
\end{figure}

\begin{figure}[!ht]
\begin{center}

\begin{tabular}{c c}
\includegraphics{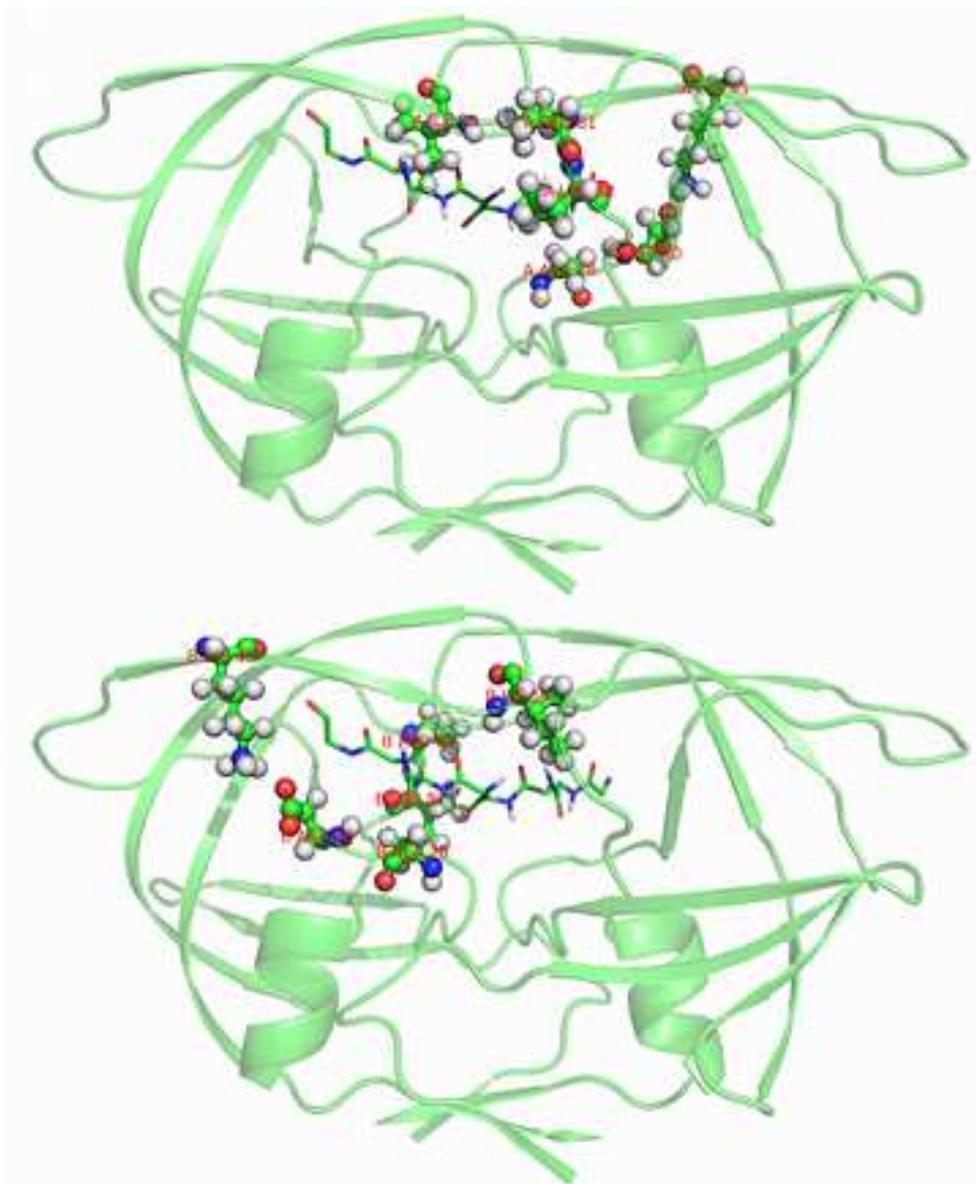}\\
\end{tabular}

\end{center}

\caption{
{\bf Spatial disposition of the residues changed by Strategy2.} The six residues of chain A (top) and 6 residues of chain B (bottom) are highlighted in ball-and-sticks. The reported structure (as semi-transparent cartoon) is that of WT-PR optimized when binding the target peptide (only the backbone is shown in sticks). Figures~\ref{mutations-chain-a-28} - \ref{mutations-chain-b-82} report the full residue by residue changes. A movie showing the three dimensional structure is included as Supporting Material.
}
\label{mutated-residues}
\end{figure}

\begin{figure}[!ht]
\begin{center}
\includegraphics{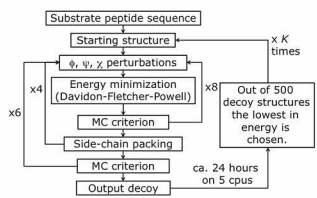}
\end{center}
\caption{
{\bf PyRosetta based optimization algorithm.} $\phi$, $\psi$, $\chi$ represent perturbations applied to both backbone and side chain dihedral angles. MC criterion stands for a Monte Carlo based check of decoy structures.
}
\label{flowchart1}
\end{figure}

\begin{figure}[!ht]
\begin{center}
\includegraphics{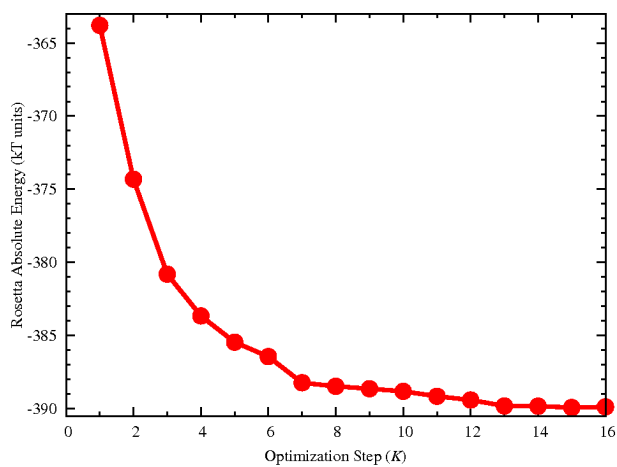}
\end{center}
\caption{
{\bf Optimization algorithm convergence.} Example of energy convergence during the various macro cycles of the optimization algorithm for WT-PR in complex with TF-PR peptide. Each point along the graph corresponds to the energy (computed with the RosettaDock energy function) of the lowest in energy decoy out of 500 produced during each of the \emph{K} steps.
}
\label{convergence}
\end{figure}

\begin{figure}[!ht]
\begin{center}
\includegraphics{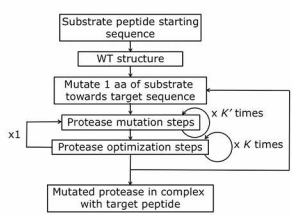}
\end{center}
\caption{
{\bf PyRosetta based mutation algorithm.} The optimization step is the algorithm presented in Figure~\ref{flowchart1}. In the mutation step the side chain perturbation for the six specificity determining residues is among all possible rotamer of all 20 amino acids.
}
\label{flowchart2}
\end{figure}

\clearpage

\section*{Tables}

\begin{table}[htp!]
\caption{
{\bf Computed binding energies of WT-PR and cleavable endogenous peptides.}
}
\begin{tabular*}{\textwidth}{@{\extracolsep{\fill}}*{4}{x{\quartpage}@{}}}
\hline
Substrate Peptide & FMO (kcal/mol) & RosettaDock (kT) & exp $K_m$ (mM)\cite{Toszer1991}  \\
\hline
MA-CA&-57 &-9& 0.15 \\
CA-p2&-52 &-6& 0.01 \\
p2-NC&-72 &-4& 0.05 \\
NC-p1&-68 &-3&  \\
p1-p6&-47 &1&  \\
p6pol-PR&-41 &-6&  \\
TF-PR&-62 &-5& $<$0.01 \\
PR-RTp51&-64 &-7& 0.07 \\
RTp51-RTp66&-68 &-12& 0.04 \\
RTp66-INT&-62 &-6&  \\
RH-IN&-63 &-10& 0.006 \\
\hline
\end{tabular*}
\label{tablebinding1}
\end{table}

\newlength\utest
\settowidth{\utest}{(only endogenous VS non-cleavable) }

\begin{table}[htp!]
\caption{
{\bf Comparison of WT-PR computed binding energies.}
}

\begin{tabular*}{\textwidth}{@{\extracolsep{\fill}} x{\utest} @{}| x{\quartpage}@{} x{\quartpage}}\hline
                          & RosettaDock Energy Function & FMO Energy Re-evaluation\\
\hline
Average endogenous$^a$    & -6  (kT)             & -60 (kcal/mol) \\
(Standard deviation)      & (3)   (kT)             & (10) (kcal/mol) \\
Average all cleavable$^b$ & -5    (kT)             & -43 (kcal/mol) \\
(Standard deviation)      & (3)   (kT)             & (22) (kcal/mol) \\
Average non cleavable$^c$ & -1   (kT)              & -15 (kcal/mol) \\
(Standard deviation)      & (4)  (kT)              & (28) (kcal/mol) \\
\hline
U test probability \mbox{(all cleavable VS non-cleavable)}   & $1.46 \cdot 10^{-7}$ & $2.21 \cdot 10^{-7}$ \\
U test probability \mbox{(only endogenous VS non-cleavable)} & $3.49 \cdot 10^{-4}$ & $6.16 \cdot 10^{-6}$ \\
\hline
\multicolumn{3}{l}{a Table~\ref{tablebinding1}.}\\
\multicolumn{3}{l}{b Table~\ref{tablebinding1} plus Table~\ref{tablebindingextra}.}\\
\multicolumn{3}{l}{c Table~\ref{tablebinding2}.}
\end{tabular*}
\begin{flushleft}
Energies were computed with the standard RosettaDock energy function, as described in \cite{Chaudhury2009} and with the FMO re-evaluation. A Mann-Whitney's U test probability was evaluated by comparing the binding energies of the set of endogenous peptide against the non-cleavable and the entire set of cleavable peptides against the non-cleavable. The FMO based binding energies are more clear in discriminating cleavable and non cleavable peptides than the Rosetta based ones.
\end{flushleft}
\label{tableenergiescomparison}
\end{table}
\begin{table}[htp!]
\caption{
{\bf FMO computed binding energies of HIV-1 protease WT and Pr3 set of mutant PRs.}
}
\begin{tabular*}{\textwidth}{@{\extracolsep{\fill}}c | c c c c c c}
\hline
PR                 & Peptides\\
             & RTp51-RTp66 & poly-Ala & poly-Arg & TF-PR & CA-p2 & p2-NC \\
\hline
WT-PR        & -68 & -15 & -41 & -62 & -52 & -72 \\
\multicolumn{1}{l |}{\emph{Single mutant}}\\
A28S & -65 & -12 & -35 & -41 & -13 &  -67 \\
D30F & -48 & -1 & 0 & -7  & -4 &  -43 \\
G48R & -66 & -44 & -27 & -55  & -15 & -43 \\
\multicolumn{1}{l |}{\emph{Double mutant}}\\
A28SD30F & -44 & -16 & 30 & -35  & -22 & -55 \\
A28SG48R & -96 & -16 & -54 & -35  & -14 & -60 \\
D30FG48R & -76 & 3 & 18 & -19  & -1 & -54 \\
\multicolumn{1}{l |}{\emph{Triple mutant}}\\
A28SD30FG48R & -42 & -21 & 12 & -32  & -12 & -63  \\
\hline
\end{tabular*}
\begin{flushleft}
Computed binding energies (kcal/mol) of WT and mutant HIV-1 proteases in complex with RTp51-RTp66, poly-alanine, poly-arginine, TF-PR, CA-p2 and p2-NC peptides.
\end{flushleft}
\label{tablebindingmutantspr3}
\end{table}
\begin{table}[htp!]
\caption{
{\bf FMO computed binding energies of HIV-1 protease WT and selected drug resistance mutant PRs (DR set).}
}
\begin{tabular*}{\textwidth}{@{\extracolsep{\fill}}c | c c c c c}
\hline
PR                 & Peptides\\
          & NC-p1$^{WT}$ & NC-p1$^{K436R}$ & NC-p1$^{A431V}$ & poly-Ala & poly-Arg \\
\hline
WT-PR           & -49 & -70 & -56 & -15 & -41 \\
D30N          & -29 & -44 & -36 & -15 & -23 \\
I50L            & -64 & -49 & -54 & -16 & -49 \\
I50V            & -54 & -46 & -52 & -18 & -34 \\
V82A           & -45 & -54 & -52 & -16 & -38 \\
I84V            & -46 & -75 & -35 & -23 & -38 \\
I84V L90M  & -55 & -67 & -55 & -20 & -46 \\
\hline
\end{tabular*}
\begin{flushleft}
Computed binding energies (kcal/mol) of WT-PR and selected drug resistance mutant proteases in complex with NC-p1 as wild type, K436R and A431V drug resistance associated mutant peptides, poly-alanine and poly-arginine peptides.
\end{flushleft}
\label{tablebindingresistance}
\end{table}
\begin{table}[htp!]
\caption{
{\bf Strategy2 suggested mutant PRs.}
}
\resizebox{\textwidth}{!}{
\begin{tabular*}{\textwidth}{@{\extracolsep{\fill}}c@{}x{\mutation}@{}x{\mutation}@{}x{\mutscheme}@{}c}
\hline
Mutant ID  & Chain A & Chain B & Mutation Scheme& Notes\\
\hline
{\bf M17} & A28S D30T         & A28S D30T K45M I50L V82F & F & After one mutation step \\
{\bf M18} & A28S D30T I50L P81D V82R I84V & A28S D30T K45M I50L V82Y & F & After one mutation step \\
{\bf M19} & A28S D30T I50L P81D V82R I84V & A28S D30T K45A I50L V82Y I84L & F & After two mutation steps \\
{\bf M20} & A28S D30T I50L P81D V82R I84V & A28S D30T K45D I50L V82Y I84L & F & After three mutation steps \\
{\bf M21} & A28S D30T I50L P81L V82Y I84V & A28S D30T K45D I50L V82Y & F & After four mutation steps \\
{\bf M22} & A28S D30T I50L P81L V82Y & A28S D30T K45A I50L V82Y & F & After five mutation steps \\
{\bf M23} & A28S D30T I50L P81L V82Y & A28S D30T K45A I50L V82Y & F & After six mutation steps \\
{\bf M24} & A28S D30T K45A I50L P81L V82Y & A28S D30T K45A I50L P81L V82Y & -- &  Homodimer of {\bf M22} \\
\hline
\end{tabular*}
}
\begin{flushleft}
{\bf M17} -- {\bf M23} represent the subsequent generations of mutant PRs suggested by Strategy2. All mutant enzymes were generated following Scheme F.
\end{flushleft}
\label{tablemutants-str2}
\end{table}
\begin{table}[htp!]
\caption{
{\bf FMO computed binding energies of HIV-1 protease WT and Strategy2 mutant PRs.}
}
\begin{tabular*}{\textwidth}{@{\extracolsep{\fill}}c | c c c c c c}
\hline
PR                 & Peptides\\
        & Target & poly-Ala & poly-Arg & TF-PR & CA-p2 & p2-NC \\
\hline
WT-PR        & -9 & -15 & -41 & -62 & -52 & -72 \\
\multicolumn{1}{l |}{\emph{Gen 1}}\\
{\bf M17}  & -34 & -13 & -47 & -68 & -82 &  -74 \\
{\bf M18}  & -24 & -19 & -45 & -82 & -62 &  -63 \\
\multicolumn{1}{l |}{\emph{Gen 2}}\\
{\bf M19}  & -17 & 2 & 1 & -67 & -46 &  -81 \\
\multicolumn{1}{l |}{\emph{Gen 3}}\\
{\bf M20}  & -23 & -2 & -19 & -67 & -33 &  -84 \\
\multicolumn{1}{l |}{\emph{Gen 4}}\\
{\bf M21}  & -20 & 2 & 7 & -37 & -32 &  -18 \\
\multicolumn{1}{l |}{\emph{Gen 5}}\\
{\bf M22}  & -29 & -6 & 10 & -42 & -25 &  -30 \\
{\bf M24}  & -29 & -11 & 7 & -44 & -33 &  -33 \\
\hline
\end{tabular*}
\begin{flushleft}
Computed binding energies (kcal/mol) of WT-PR and Strategy2 mutant proteases in complex with Target, poly-alanine, poly-arginine, TF-PR, CA-p2 and p2-NC peptides.
\end{flushleft}
\label{tablebindingmutantstr3}
\end{table}

\clearpage

\newpage

\section*{Supporting Material for}
\vspace{0.5cm}
{\LARGE\bf\emph{In silico} prediction of mutant HIV-1 proteases  cleaving a target sequence}

\vspace{0.5cm}
\noi Jan H. Jensen,
Martin Willemo{\"{e}}s,
Jakob R. Winther,
Luca De Vico
\vspace{0.5cm}

\setcounter{page}{1}
\renewcommand{\thepage}{S-\arabic{page}}

\markboth{}{}

The video animation of the optimized structure of WT-PR binding the target peptide with highlighted residues can be found at this link: \href{http://youtu.be/NEXKojTw2Bc}{http://youtu.be/NEXKojTw2Bc} .

\begin{tablesupp}[h]
\caption{
{\bf Cleavable peptides.}
}
\begin{tabular}{lccccccccc}
\hline
            & P4  & P3  & P2  & P1  & * & P1' & P2' & P3' & P4' \\
\hline
MA-CA&S&Q&N&Y&*&P&I&V&Q\\
CA-p2&A&R&V&L&*&A&E&A&M\\
p2-NC&A&T&I&M&*&M&Q&R&G\\
NC-p1&R&Q&A&N&*&F&L&G&K\\
p1-p6&P&G&N&F&*&L&Q&S&R\\
p6pol-PR&S&F&N&F&*&P&Q&V&T\\
TF-PR&S&F&N&F&*&P&Q&I&T\\
PR-RTp51&T&L&N&F&*&P&I&S&P\\
RTp51-RTp66&A&E&T&F&*&Y&V&D&G\\
RTp66-INT&R&K&V&L&*&F&L&D&G\\
RH-IN&R&K&I&L&*&F&L&D&G\\
\hline
\end{tabular}
\begin{flushleft}
List of the cleavable endogenous peptides considered in this work
\end{flushleft}
\label{tablepeptides1}
\end{tablesupp}

\newlength{\ncth}
\setlength{\ncth}{\textheight}
\addtolength{\ncth}{-12cm}

\begin{tablesupp}
\caption{
{\bf Extra cleavable peptides.}
}
\resizebox{!}{\ncth}{
\begin{tabular}{lccccccccc}
\hline
            & P4  & P3  & P2  & P1  & * & P1' & P2' & P3' & P4' \\
\hline
K001 & T & Q & I & M & * & F & E & T & F  \\
K002 & G & Q & V & N & * & Y & E & E & F  \\
K003 & P & F & I & F & * & E & E & E & P  \\
K005 & D & T & V & L & * & E & E & M & S  \\
K007 & A & E & E & L & * & A & E & I & F  \\
K008 & S & L & N & L & * & R & E & T & Q  \\
K010 & A & E & C & F & * & R & I & F & D  \\
K011 & D & Q & I & L & * & I & E & I & C  \\
K012 & D & D & L & F & * & F & E & A & D  \\
K013 & Y & E & E & F & * & V & Q & M & M  \\
K014 & P & I & V & G & * & A & E & T & F  \\
K016 & R & E & A & F & * & R & V & F & D  \\
K018 & A & Q & T & F & * & Y & V & N & L  \\
K019 & P & T & L & L & * & T & E & A & P  \\
K020 & S & F & I & G & * & M & E & F & K  \\
K021 & D & A & I & N & * & T & E & F & K  \\
K022 & Q & I & T & L & * & W & Q & R & P  \\
K023 & E & L & E & F & * & P & E & G & G  \\
K029 & K & E & L & Y & * & P & L & T & S  \\
K031 & S & R & S & L & * & Y & A & S & S  \\
K032 & A & E & A & M & * & S & Q & V & T  \\
K034 & G & S & H & L & * & V & E & A & L  \\
K035 & G & G & V & Y & * & A & T & R & S  \\
K036 & F & R & S & G & * & V & E & T & T  \\
K037 & V & E & V & A & * & E & E & E & E  \\
K038 & L & P & V & N & * & G & E & F & S  \\
K039 & E & T & T & A & * & L & V & C & D  \\
K040 & H & L & V & E & * & A & L & Y & L  \\
K041 & H & Y & G & F & * & P & T & Y & G  \\
K042 & D & S & A & D & * & A & E & E & D  \\
K043 & G & W & I & L & * & G & E & H & G  \\
K045 & Q & A & I & Y & * & L & A & L & Q  \\
K046 & E & K & V & Y & * & L & A & W & V  \\
K047 & V & E & I & C & * & T & E & M & E  \\
K048 & T & Q & D & F & * & W & E & V & Q  \\
K049 & L & W & M & G & * & Y & E & L & H  \\
K050 & G & D & A & Y & * & F & S & V & P  \\
K051 & E & L & E & L & * & A & E & N & R  \\
K052 & S & K & D & L & * & I & A & E & I  \\
K053 & L & E & V & N & * & I & V & T & D  \\
K054 & I & I & V & A & * & C & E & G & N  \\
K056 & G & G & N & Y & * & P & V & Q & H  \\
K057 & A & R & L & M & * & A & E & A & L  \\
K058 & P & F & A & A & * & A & Q & Q & R  \\
K059 & P & R & N & F & * & P & V & A & Q  \\
K060 & G & L & A & A & * & P & Q & F & S  \\
K061 & S & L & N & L & * & P & V & A & K  \\
K063 & R & Q & V & L & * & F & L & E & K  \\
K064 & Q & M & I & F & * & E & E & H & G  \\
SUB3 & Q & I & T & L & * & W & K & R & P  \\
T035 & V & E & I & C & * & T & E & M & E  \\
T084 & T & Q & D & F & * & W & E & V & Q  \\
T112 & G & D & A & Y & * & F & S & V & P  \\
T228 & L & W & M & G & * & Y & E & L & H  \\
T300 & E & L & E & L & * & A & E & N & R  \\
T322 & S & K & D & L & * & I & A & E & I  \\
T480 & Q & A & I & Y & * & L & A & L & Q  \\
T491 & L & E & V & N & * & I & V & T & D  \\
T529 & E & K & V & Y & * & L & A & W & V  \\
\hline
\end{tabular}
}
\begin{flushleft}
List of the cleavable non-endogenous peptides considered in this work
\end{flushleft}
\label{tableextrapeptides}
\end{tablesupp}

\begin{tablesupp}
\caption{
{\bf Other peptides}
}
\begin{tabular}{lccccccccc}
\hline
            & P4  & P3  & P2  & P1  & * & P1' & P2' & P3' & P4' \\
\hline
Target&H&L&S&F&*&M&A&I&P\\
NC-p1$^{A431V}$&R&Q&V&N&*&F&L&G&K\\
NC-p1$^{K436R}$&R&Q&A&N&*&F&L&G&R\\
poli-Ala&A&A&A&A&*&A&A&A&A\\
poli-Arg&R&R&R&R&*&R&R&R&R\\
\hline
\end{tabular}
\begin{flushleft}
List of other peptides considered in this work
\end{flushleft}
\label{tablepeptides2}
\end{tablesupp}

\begin{tablesupp}
\caption{
{\bf Non-cleavable peptides}
}
\resizebox{!}{\ncth}{
\begin{tabular}{lccccccccc}
\hline
            & P4  & P3  & P2  & P1  & * & P1' & P2' & P3' & P4' \\
\hline
NBP1&V&N&C&A&*&K&K&I&V\\
NBP2&W&R&N&R&*&C&K&G&T\\
NBP3&M&M&K&S&*&R&N&L&T\\
NBP4&L&A&A&A&*&M&K&R&H\\
NBP5&T&T&Q&A&*&N&K&H&I\\
T015&G&M&D&G&*&P&K&V&K\\
T031&I&K&A&L&*&V&E&I&C\\
T033&A&L&V&E&*&I&C&T&E\\
T037&I&C&T&E&*&M&E&K&E\\
T039&T&E&M&E&*&K&E&G&K\\
T080&L&N&K&R&*&T&Q&D&F\\
T082&K&R&T&Q&*&D&F&W&E\\
T086&D&F&W&E&*&V&Q&L&G\\
T088&W&E&V&Q&*&L&G&I&P\\
T108&V&L&D&V&*&G&D&A&Y\\
T110&D&V&G&D&*&A&Y&F&S\\
T114&A&Y&F&S&*&V&P&L&D\\
T116&F&S&V&P&*&L&D&E&D\\
T224&E&P&P&F&*&L&W&M&G\\
T226&P&F&L&W&*&M&G&Y&E\\
T230&M&G&Y&E&*&L&H&P&D\\
T232&Y&E&L&H&*&P&D&K&W\\
T296&T&E&E&A&*&E&L&E&L\\
T298&E&A&E&L&*&E&L&A&E\\
T302&E&L&A&E&*&N&R&E&I\\
T304&A&E&N&R&*&E&I&L&K\\
T318&Y&Y&D&P&*&S&K&D&L\\
T320&D&P&S&K&*&D&L&I&A\\
T324&D&L&I&A&*&E&I&Q&K\\
T326&I&A&E&I&*&Q&K&Q&G\\
T441&Y&V&D&G&*&A&A&N&R\\
T476&K&T&E&L&*&Q&A&I&Y\\
T478&E&L&Q&A&*&I&Y&L&A\\
T482&I&Y&L&A&*&L&Q&D&S\\
T484&L&A&L&Q&*&D&S&G&L\\
T487&Q&D&S&G&*&L&E&V&N\\
T489&S&G&L&E&*&V&N&I&V\\
T493&V&N&I&V&*&T&D&S&Q\\
T495&I&V&T&D&*&S&Q&Y&A\\
T525&L&I&K&K&*&E&K&L&A\\
T527&K&K&E&K&*&V&Y&L&A\\
T531&V&Y&L&A&*&W&V&P&A\\
T533&L&A&W&V&*&P&A&H&K\\
\hline
\end{tabular}
}
\begin{flushleft}
List of non-cleavable peptides considered in this work
\end{flushleft}
\label{tablepeptides3}
\end{tablesupp}

\begin{tablesupp}[htp!]
\caption{
{\bf Computed binding energies of WT-PR and non-endogenous cleavable peptides.}
}
\begin{tabular*}{\textwidth}{@{\extracolsep{\fill}}x{\sixthpage}@{}x{\sixthpage}@{}x{\sixthpage}|x{\sixthpage}@{}x{\sixthpage}@{}x{\sixthpage}}
\hline
Substrate Peptide & FMO (kcal/mol) & RosettaDock (kT) & Substrate Peptide & FMO (kcal/mol) & RosettaDock (kT) \\
\hline
K001  &  -22  &-2&  K043  &  -67  &-5\\
K002  &  -33  &-4&  K045  &  -55  &-5\\
K003  &  -16  &-8&  K046  &  -39  &-4\\
K005  &  -24  &-4&  K047  &  -79  &-5\\
K007  &  -27  &-1&  K048  &  -86  &-9\\
K008  &  -40  &-2&  K049  &  -40  &-5\\
K010  &  -55  &-4&  K050  &  -36  &-7\\
K011  &  -42  &-3&  K051  &  -31  &-2\\
K012  &  -22  &-7&  K052  &  -4  &1\\
K013  &  -36  &-2&  K053  &  -51  &-4\\
K014  &  -29  &-3&  K054  &  -56  &-1\\
K016  &  -73  &-3&  K056  &  -70  &-5\\
K018  &  -72  &-4&  K057  &  -7  &-6\\
K019  &  -61  &-9&  K058  &  -35  &-2\\
K020  &  -8  &-3&  K059  &  -58  &-6\\
K021  &  -64  &-8&  K060  &  -36  &-5\\
K022  &  -45  &-5&  K061  &  -34  &-5\\
K023  &  -63  &-11&  K063  &  -61  &-3\\
K029  &  -48  &-4&  K064  &  -2  &-7\\
K031  &  -32  &-5&  SUB3  &  -30  &-6\\
K032  &  -51  &-6&  T035  &  -67  &-4\\
K034  &  -8  &-8&  T084  &  -81  &-7\\
K035  &  -30  &-8&  T112  &  -69  &-11\\
K036  &  -3  &-3&  T228  &  -29  &-5\\
K037  &  -2  &-4&  T300  &  -29  &-1\\
K038  &  -39  &-5&  T322  &  -21  &1\\
K039  &  -10  &0&  T480  &  -62  &-5\\
K040  &  -30  &-6&  T491  &  -56  &-1\\
K041  &  -53  &-6&  T529  &  -27  &1\\
K042  &  -29  &-6 \\
\hline
\end{tabular*}
\label{tablebindingextra}
\end{tablesupp}

\begin{tablesupp}[htp!]
\caption{
{\bf Computed binding energies of WT-PR and non-cleavable peptides.}
}
\begin{tabular*}{\textwidth}{@{\extracolsep{\fill}}x{\sixthpage}@{}x{\sixthpage}@{}x{\sixthpage}|x{\sixthpage}@{}x{\sixthpage}@{}x{\sixthpage}}
\hline
Substrate Peptide & FMO (kcal/mol) & RosettaDock (kT) & Substrate Peptide & FMO (kcal/mol) & RosettaDock (kT) \\
\hline
NBP1&-18&3&T296&43&2\\
NBP2&-21&3&T298&-54&-2\\
NBP3&-63&0&T302&-44&4\\
NBP4&-18&4&T304&1&-2\\
NBP5&-68&7&T318&23&0\\
T015&2&-3&T320&23&-2\\
T031&-29&-6&T324&-18&0\\
T033&-10&-2&T326&-10&5\\
T037&-28&-5&T441&-30&-2\\
T039&-36&-3&T476&-10&1\\
T080&-45&4&T478&-33&-2\\
T082&9&-2&T482&-45&-1\\
T086&-23&0&T484&-24&-2\\
T088&-42&3&T487&31&-3\\
T108&-12&-4&T489&-24&-6\\
T110&-21&-3&T493&-19&-1\\
T114&-16&-8&T495&-15&1\\
T116&-10&-2&T525&61&7\\
T224&48&-1&T527&-9&1\\
T226&-49&-9&T531&-5&-5\\
T230&19&-2&T533&-23&-7\\
T232&-13&-1\\
\hline
\end{tabular*}
\label{tablebinding2}
\end{tablesupp}

\begin{tablesupp}[htp!]
\caption{
{\bf Strategy1 suggested mutant PRs.}
}
\resizebox{\textwidth}{!}{
\begin{tabular*}{\textwidth}{@{\extracolsep{\fill}}c@{}x{\mutation}@{}x{\mutation}@{}x{\mutscheme}@{}c}
\hline
Mutant ID  & Chain A & Chain B & Mutation Scheme & Notes\\
\hline
{\bf M1} & V82R I84V & D30Y V82I & A & \\
{\bf M2} & V82Y          & D30V         & B & \\
{\bf M3} & D30T          & D30V V82I & B F& \\
{\bf M4} & D30Y V82R & D30Y V82R & -- & Homodimer of {\bf M1}\\
{\bf M5} & D30V V82Y & D30V V82Y & -- & Homodimer of {\bf M2}\\
{\bf M6} & D30V           & D30V           & -- & Homodimer of {\bf M3}\\
{\bf M7} & D30T I84V & D30V V82F & A F & Initial temperature = 9 kT\\
{\bf M8} &                    & D30V          & B F & Initial temperature = 9 kT\\
{\bf M9} & D30T I47L L76F V82R I84T & D30E V82Y & A & Initial temperature = 6 kT\\
{\bf M10} & V82Y & D30T I84L & B & Initial temperature = 6 kT\\
{\bf M11} & D30V V82Y I84V & D30H I47L L76F V82Y & A & Initial temperature = 12 kT\\
{\bf M12} & V82Y & D30T & B & Initial temperature = 12 kT\\
{\bf M13} & D30E L76F V82R & D30E L76F V82R & -- & Homodimer of {\bf M9}\\
{\bf M14} & D30T V82Y I84L & D30T V82Y I84L & -- & Homodimer of {\bf M10}\\
{\bf M15} & D30H I47L V82Y & D30H I47L V82Y & -- & Homodimer of {\bf M11}\\
{\bf M16} & D30T V82Y & D30T V82Y & -- & Homodimer of {\bf M12}\\
\hline
\end{tabular*}
}
\begin{flushleft}
Different mutant PRs were obtained by small modifications of the mutation algorithm. Inside Strategy1 two different schemes were used when choosing which residues could mutate. In Scheme A all six specificity determining residues were allowed to mutate on both chains. In Scheme B only residues 76 and 82 were set as mutable on Chain A and 30, 47, 48, and 84 on Chain B. In addition, a straight forward variant of the algorithm was tested, as opposed to the step-wise one presented in Table~\ref{tablemutation}. In this variant (Scheme F) the protease was directly 'exposed' to the final target peptide sequence. A $K$ value in the order of 20 was necessary. Other parameters that differ from those specified in the Computational Methods section are also highlighted.
\end{flushleft}
\label{tablemutants-str1}
\end{tablesupp}

\begin{tablesupp}[htp!]
\caption{
{\bf FMO computed binding energies of HIV-1 protease WT and Strategy1 mutant PRs.}
}
\begin{tabular*}{\textwidth}{@{\extracolsep{\fill}}c | c c c c c c}
\hline
PR                 & Peptides\\
        & Target & poly-Ala & poly-Arg & TF-PR & CA-p2 & p2-NC \\
\hline
WT-PR        & -9 & -15 & -41 & -62 & -52 & -72 \\
{\bf M1} & -7 & 3 & 28 & -41 \\
{\bf M2} & -13 & -9 & 12 & -41 \\
{\bf M3} & -8 & -4 & 13 & -43 \\
{\bf M4} & -18 & -24 & 17 & -52 \\
{\bf M5} & -30 & -14 & 4 & -54 & -27 & -73 \\
{\bf M6} & -14 & 2 & -1 & -14 \\
{\bf M7} & -7   & -2 & 3 & -49 \\
{\bf M8} & -10 & -9 & -5 & -55 \\
{\bf M9} & 20 & 17 & 14 & -36 \\
{\bf M10} & 40 & 11 & -3 & -43 \\
{\bf M11} & 5 & 10 & 45 & -44 \\
{\bf M12} & 2 & 2 & 16 & -43 \\
{\bf M13} & 3 & 2 & -24 & -52 \\
{\bf M14} & 3 & 8 & 17 & -60 \\
{\bf M15} & -9 & 2 & 71 & -51 \\
{\bf M16} & -8 & -6 & 26 & -44 \\
\hline
\end{tabular*}
\begin{flushleft}
Computed binding energies (kcal/mol) of WT-PR and Strategy1 mutant proteases in complex with target, poly-alanine, poly-arginine, TF-PR, CA-p2 and p2-NC peptides.
\end{flushleft}
\label{tablebindingmutants}
\end{tablesupp}

\begin{tablesupp}[htp!]
\caption{
{\bf Residues set as mutable in Strategy 2.}
}
\begin{tabular}{c c}
\hline
Chain A & Chain B \\
\hline
Arg 8 & Arg 8 \\
Ala 28 & Leu 23 \\
Asp 29 & Ala 28 \\
Asp 30 & Asp 29 \\
Val 32 & Asp 30 \\
Gly 48 & Lys 45 \\
Gly 49 & Ile 47 \\
Ile 50 & Gly 48 \\
Leu 76 & Gly 49 \\
Thr 80 & Ile 50 \\
Pro 81 & Pro 81 \\
Val 82 & Val 82 \\
Ile 84 & Ile 84 \\
\hline
\end{tabular}
\begin{flushleft}
The residues were selected as those inside a 3 {\AA} radius from the substrate peptide plus the specificity determining residues, if not included, minus the catalytic triad Asp25, Thr26 and Gly27 on both chains. The optimized structure of WT protease in complex with the TF-PR peptide was used as template.
\end{flushleft}
\label{tablemutstr2}
\end{tablesupp}

\begin{tablesupp}[!ht]
\caption{
{\bf Substrate peptide mutation sequence}
}
\begin{tabular}{lccccccccc}
\hline
            & P4  & P3  & P2  & P1  & * & P1' & P2' & P3' & P4' \\
\hline
Start   & Ser & Phe & Asn & Phe & * & Pro & Gln & Ile & Thr \\ 
           & His & Phe & Asn & Phe & * & Pro & Gln & Ile & Thr \\
           & His & Phe & Asn & Phe & * & Pro & Gln & Ile & Pro \\
           & His & Leu & Asn & Phe & * & Pro & Gln & Ile & Pro \\
           & His & Leu & Asn & Phe & * & Pro & Ala & Ile & Pro \\
           & His & Leu & Ser & Phe & * & Pro & Ala & Ile & Pro \\
Target & His & Leu & Ser & Phe & * & Met & Ala & Ile & Pro\\
\hline
\end{tabular}
\begin{flushleft}
Step wise sequence of substrate peptides employed in the mutation algorithm. The starting sequence corresponds to the natural substrate TF-PR. This sequence is altered one amino acid at the time towards that of the desired target sequence. The P1 and P3' position were not changed during the sequence.
\end{flushleft}
\label{tablemutation}
\end{tablesupp}

\newlength\eigthpage
\addtolength{\eigthpage}{\textwidth}
\addtolength{\eigthpage}{-2cm}
\eigthpage=0.125\eigthpage

\newlength{\rocth}
\setlength{\rocth}{\textheight}
\addtolength{\rocth}{-12.5cm}

\begin{tablesupp}[htp!]
\caption{
{\bf ROC data.}
}
\resizebox{!}{\rocth}{
\begin{tabular*}{\textwidth}{@{\extracolsep{\fill}}x{\eigthpage}@{}x{\eigthpage}@{}x{\eigthpage}@{}x{\eigthpage}|x{\eigthpage}@{}x{\eigthpage}@{}x{\eigthpage}@{}x{\eigthpage}}
\hline
\multicolumn{8}{c}{\emph{Total}}\\
FMO&&&&Rosetta&&&\\
&&&&&&&\\
cutoff&True positive&False positive&Distance to (0, 1)&cutoff&True positive&False positive&Distance to (0, 1)\\
 -inf&1.00&1.00&1.00& -inf&1.00&1.00&1.00\\
-10&0.90&0.67&0.68&4&1.00&0.88&0.88\\
-15&0.89&0.56&0.57&3&1.00&0.84&0.84\\
-20&0.87&0.44&0.46&2&1.00&0.77&0.77\\
-25&0.81&0.30&0.35&1&0.99&0.74&0.74\\
-30&0.71&0.26&0.38&0&0.94&0.63&0.63\\
-35&0.61&0.21&0.44&-1&0.93&0.58&0.59\\
-40&0.53&0.19&0.51&-2&0.84&0.37&0.40\\
-45&0.47&0.09&0.54&-3&0.74&0.23&0.35\\
-50&0.44&0.07&0.56&-4&0.61&0.19&0.43\\
-55&0.39&0.05&0.62&-5&0.50&0.14&0.52\\
-60&0.30&0.05&0.70&-6&0.27&0.09&0.73\\
-65&0.17&0.02&0.83&-7&0.20&0.05&0.80\\
-70&0.09&0.00&0.91&-8&0.13&0.05&0.87\\
-75&0.04&0.00&0.96&-9&0.06&0.00&0.94\\
-80&0.03&0.00&0.97&-10&0.04&0.00&0.96\\
-85&0.01&0.00&0.99&-11&0.01&0.00&0.99\\
 +inf&0.00&0.00&1.00& +inf&0.00&0.00&1.00\\
\hline
\multicolumn{8}{c}{\emph{Only endogenous}}\\
FMO&&&&Rosetta&&&\\
&&&&&&&\\
cutoff&True positive&False positive&Distance to (0, 1)&cutoff&True positive&False positive&Distance to (0, 1)\\
 -inf&1.00&1.00&1.00& -inf&1.00&1.00&1.00\\
-10&1.00&0.67&0.67&4&1.00&0.88&0.88\\
-15&1.00&0.56&0.56&3&1.00&0.84&0.84\\
-20&1.00&0.44&0.44&2&1.00&0.77&0.77\\
-25&1.00&0.30&0.30&1&0.91&0.74&0.75\\
-30&1.00&0.26&0.26&0&0.91&0.63&0.63\\
-35&1.00&0.21&0.21&-1&0.91&0.58&0.59\\
-40&1.00&0.19&0.19&-2&0.91&0.37&0.38\\
-45&0.91&0.09&0.13&-3&0.91&0.23&0.25\\
-50&0.82&0.07&0.19&-4&0.73&0.19&0.33\\
-55&0.73&0.05&0.28&-5&0.73&0.14&0.31\\
-60&0.64&0.05&0.37&-6&0.36&0.09&0.64\\
-65&0.27&0.02&0.73&-7&0.36&0.05&0.64\\
-70&0.09&0.00&0.91&-8&0.27&0.05&0.73\\
-75&0.00&0.00&1.00&-9&0.18&0.00&0.82\\
-80&0.00&0.00&1.00&-10&0.09&0.00&0.91\\
-85&0.00&0.00&1.00&-11&0.09&0.00&0.91\\
+inf&0.00&0.00&1.00& +inf&0.00&0.00&1.00\\
\hline
\end{tabular*}
}
\begin{flushleft}
Comparison of ROC data for FMO energy re-evaluation and RosettaDock energy function generated binding energies, while considering different cutoff values. The upper part of the table reports the full comparison between known cleavable and non cleavable peptides. In the lower part, data for only the endogenous peptides was used for the cleavable part. True positive data is reported in the graphs as sensitivity, false positive as 1 - specificity. Theoretical values for $\pm$ infinite cutoff have been added.
\end{flushleft}
\label{tableROCdata}
\end{tablesupp}

\begin{figuresupp}[!ht]
\begin{center}
\includegraphics{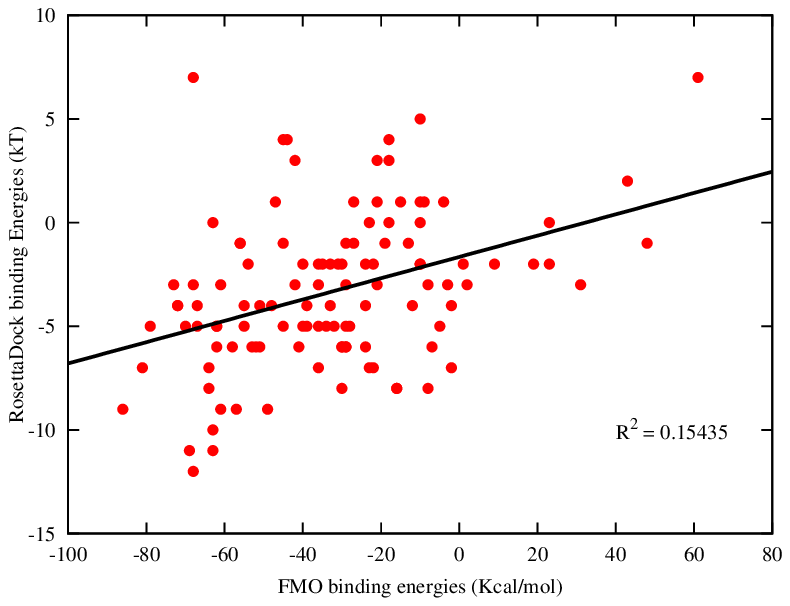}
\end{center}
\caption{
{\bf Correlation plot between FMO and RosettaDock computed binding energies.} The linear trend line shows no correlation between the data (R$^2$ = 0.15435).
}
\label{correlation}
\end{figuresupp}

\begin{figuresupp}[!ht]
\begin{center}
\includegraphics{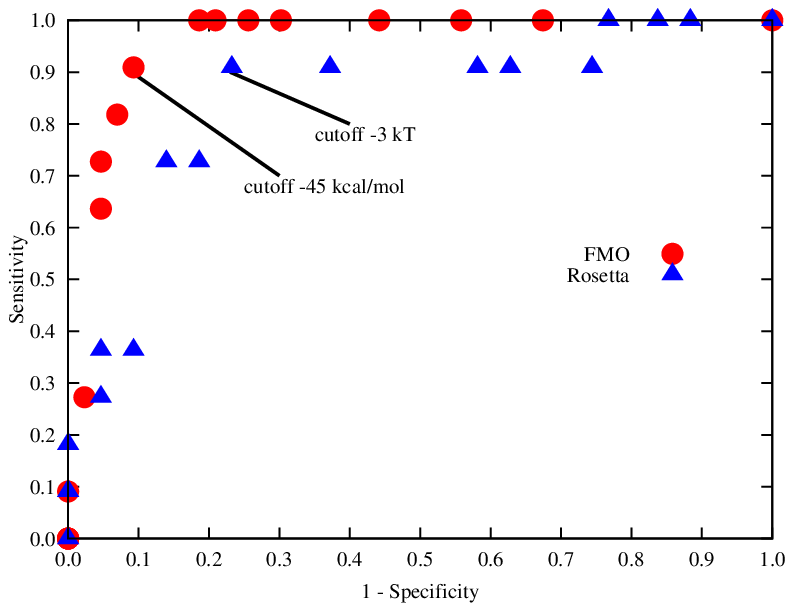}
\end{center}
\caption{
{\bf ROC plot comparing different cutoff values for binding energies computed through FMO energy re-evaluation or RosettaDock energy function.} The values for each method closest to the theoretical optimum (0,1) are highlighted. The comparison was done using the data of only the endogenous peptides for the cleavable peptides part. The computed area under the ROC curve is 0.96 and 0.84 for FMO and Rosetta, respectively. The raw data is reported in Table~\ref{tableROCdata}.
}
\label{ROCendogenous}
\end{figuresupp}

\clearpage

Figures~\ref{mutations-chain-a-28} - \ref{mutations-chain-b-82} compare the changes between WT-PR and {\bf M24 }, residue by residue. In each figure the enzyme is represented as semi-transparent ribbon, the peptide as sticks and the changing residue as ball-and-sticks. The peptide residues numbering is from 2 to 9. Each residue changed by Strategy2 is indicated by a label containing the chain, the residue name and its number.

\begin{figuresupp}

\begin{tabular}{c c}
\multicolumn{2}{c}{A28S}\\
\hline
WT  &
\includegraphics[width=12.0cm]{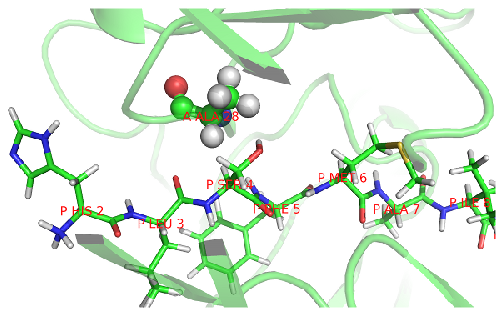} \\
{\bf M24 } &
\includegraphics[width=12.0cm]{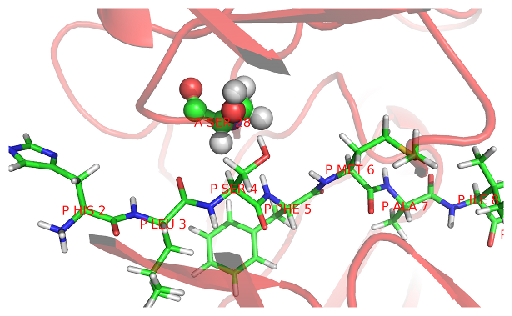} \\
\end{tabular}

\caption{
{\bf Chain A, residue 28.}
}
\label{mutations-chain-a-28}
\end{figuresupp}

\begin{figuresupp}

\begin{tabular}{c c}
\multicolumn{2}{c}{D30T}\\
\hline
WT  &
\includegraphics[width=12.0cm]{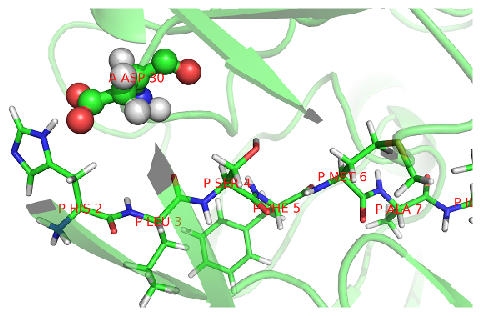} \\
{\bf M24 } &
\includegraphics[width=12.0cm]{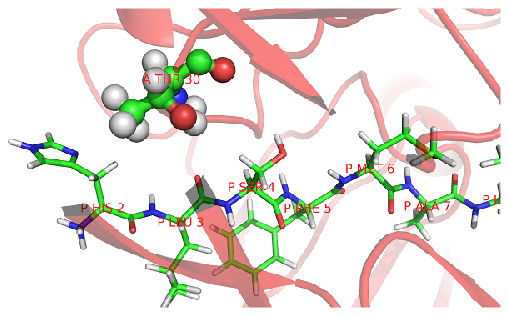} \\
\end{tabular}

\caption{
{\bf Chain A, residue 30.}
}
\label{mutations-chain-a-30}
\end{figuresupp}

\begin{figuresupp}

\begin{tabular}{c c}
\multicolumn{2}{c}{K45A}\\
\hline
WT  &
\includegraphics[width=12.0cm]{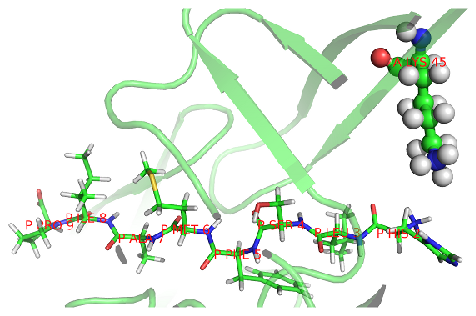} \\
{\bf M24 } &
\includegraphics[width=12.0cm]{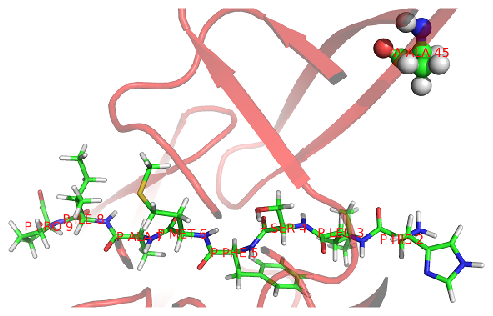} \\
\end{tabular}

\caption{
{\bf Chain A, residue 45.}
}
\label{mutations-chain-a-45}
\end{figuresupp}

\begin{figuresupp}

\begin{tabular}{c c}
\multicolumn{2}{c}{I50L}\\
\hline
WT  &
\includegraphics[width=12.0cm]{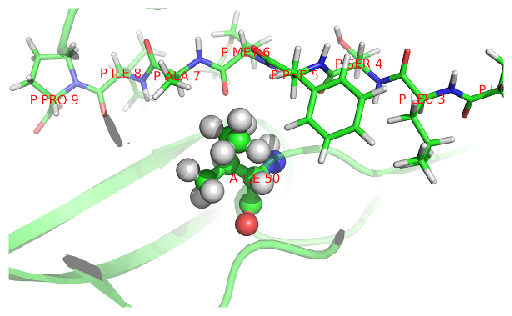} \\
{\bf M24 } &
\includegraphics[width=12.0cm]{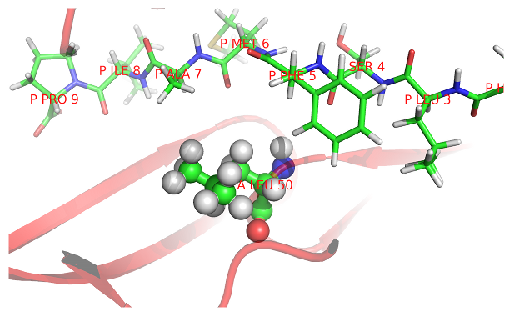} \\
\end{tabular}

\caption{
{\bf Chain A, residue 50.}
}
\label{mutations-chain-a-50}
\end{figuresupp}

\begin{figuresupp}

\begin{tabular}{c c}
\multicolumn{2}{c}{P81L}\\
\hline
WT  &
\includegraphics[width=12.0cm]{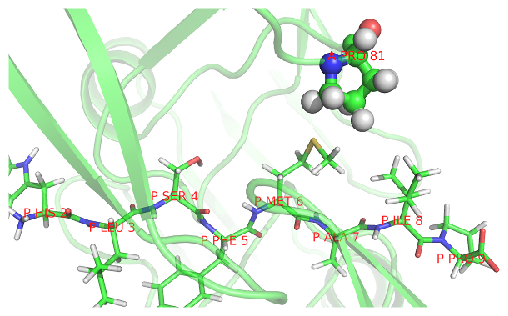} \\
{\bf M24 } &
\includegraphics[width=12.0cm]{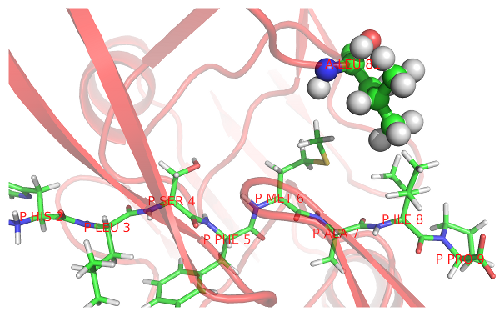} \\
\end{tabular}

\caption{
{\bf Chain A, residue 81.}
}
\label{mutations-chain-a-81}
\end{figuresupp}

\begin{figuresupp}

\begin{tabular}{c c}
\multicolumn{2}{c}{V82Y}\\
\hline
WT  &
\includegraphics[width=12.0cm]{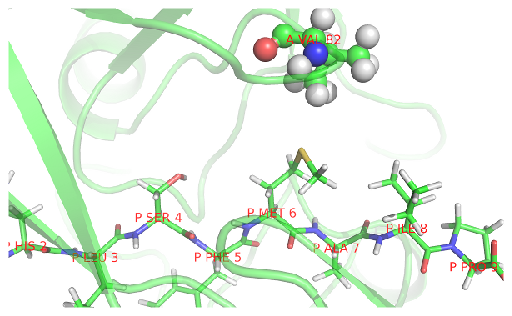} \\
{\bf M24 } &
\includegraphics[width=12.0cm]{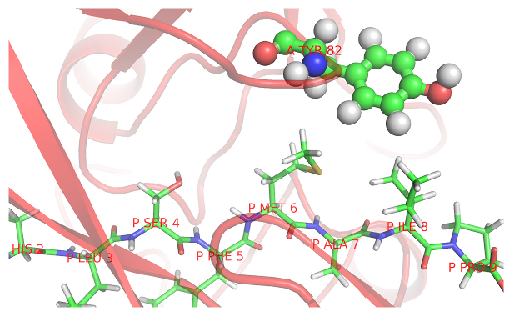} \\
\end{tabular}

\caption{
{\bf Chain A, residue 82.}
}
\label{mutations-chain-a-82}
\end{figuresupp}

\clearpage

\begin{figuresupp}

\begin{tabular}{c c}
\multicolumn{2}{c}{A28S}\\
\hline
WT  &
\includegraphics[width=12.0cm]{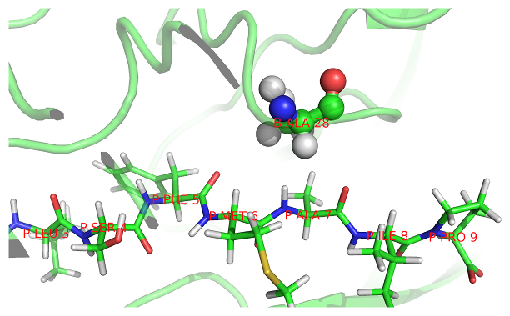} \\
{\bf M24 } &
\includegraphics[width=12.0cm]{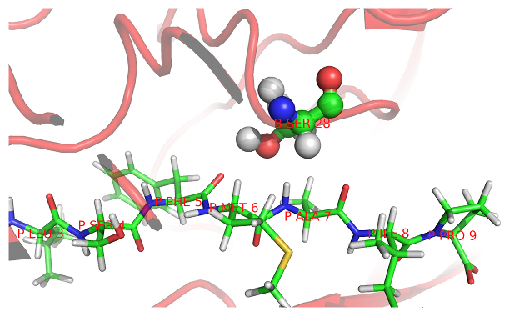} \\
\end{tabular}

\caption{
{\bf Chain B, residue 28.}
}
\label{mutations-chain-b-28}
\end{figuresupp}

\begin{figuresupp}

\begin{tabular}{c c}
\multicolumn{2}{c}{D30T}\\
\hline
WT  &
\includegraphics[width=12.0cm]{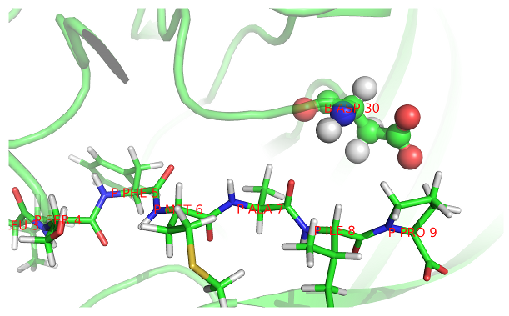} \\
{\bf M24 } &
\includegraphics[width=12.0cm]{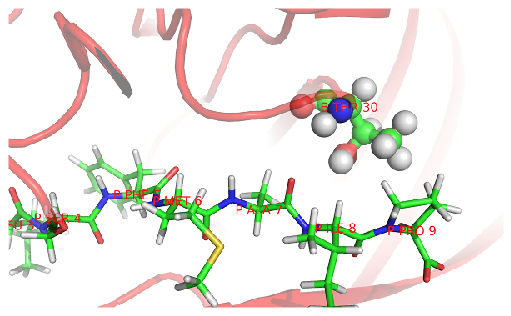} \\
\end{tabular}

\caption{
{\bf Chain B, residue 30.}
}
\label{mutations-chain-b-30}
\end{figuresupp}

\begin{figuresupp}

\begin{tabular}{c c}
\multicolumn{2}{c}{K45A}\\
\hline
WT  &
\includegraphics[width=12.0cm]{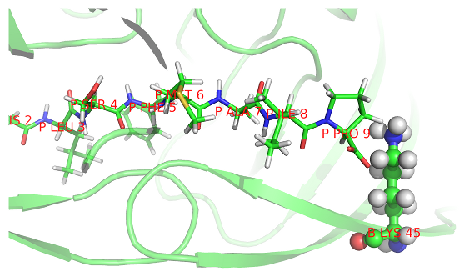} \\
{\bf M24 } &
\includegraphics[width=12.0cm]{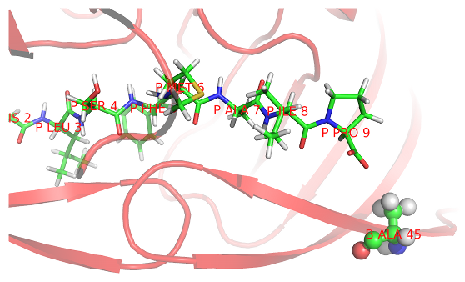} \\
\end{tabular}

\caption{
{\bf Chain B, residue 45.}
}
\label{mutations-chain-b-45}
\end{figuresupp}

\begin{figuresupp}

\begin{tabular}{c c}
\multicolumn{2}{c}{I50L}\\
\hline
WT  &
\includegraphics[width=12.0cm]{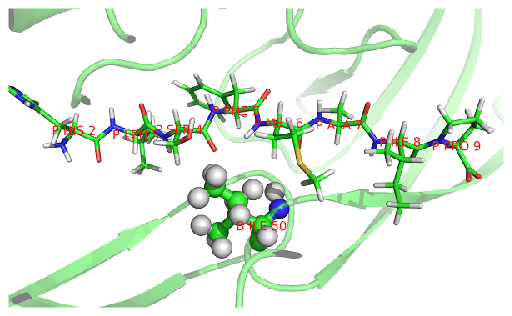} \\
{\bf M24 } &
\includegraphics[width=12.0cm]{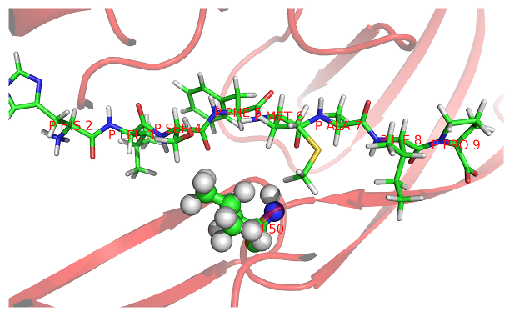} \\
\end{tabular}

\caption{
{\bf Chain B, residue 50.}
}
\label{mutations-chain-b-50}
\end{figuresupp}

\begin{figuresupp}

\begin{tabular}{c c}
\multicolumn{2}{c}{P81L}\\
\hline
WT  &
\includegraphics[width=12.0cm]{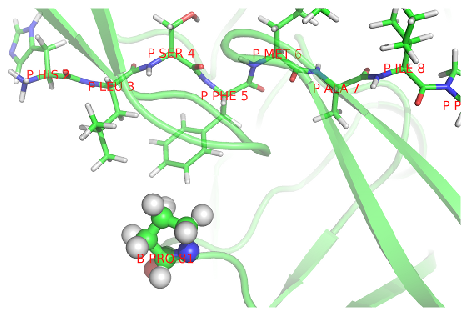} \\
{\bf M24 } &
\includegraphics[width=12.0cm]{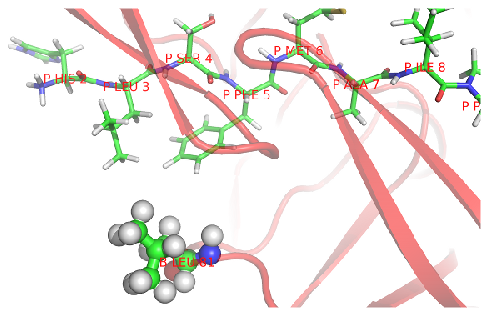} \\
\end{tabular}

\caption{
{\bf Chain B, residue 81.}
}
\label{mutations-chain-b-81}
\end{figuresupp}

\begin{figuresupp}

\begin{tabular}{c c}
\multicolumn{2}{c}{V82Y}\\
\hline
WT  &
\includegraphics[width=12.0cm]{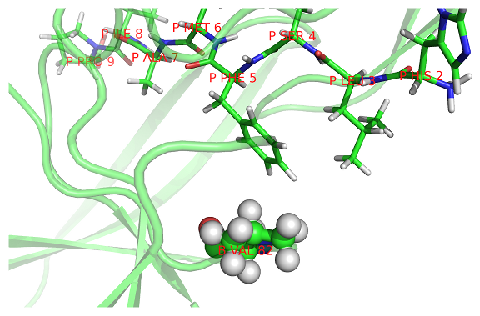} \\
{\bf M24 } &
\includegraphics[width=12.0cm]{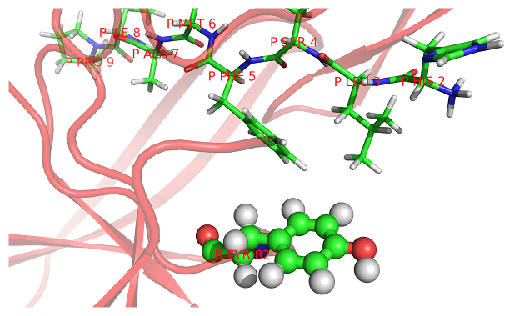} \\
\end{tabular}

\caption{
{\bf Chain B, residue 82.}
}
\label{mutations-chain-b-82}
\end{figuresupp}

\end{document}